\documentclass[preprint,a4paper,12pt,3p]{elsarticle}
\usepackage{epsf,amsmath,amssymb,graphicx,scalefnt}
\usepackage[usenames,dvipsnames]{color}
\usepackage{listings}
\biboptions{numbers,sort&compress}
\newcommand{\Q}{{\scriptstyle Q}}
\newcommand{\qqh}{\Q'\bar \Q\phi}
\newcommand{\rge}{{\abbrev RGE}}
\newcommand{\NS}{N}
\newcommand{\NM}{P}
\newcommand{\X}{{\abbrev X}}
\newcommand{\blockentry}[3]{{\tt #1(#2)#3}}
\newcommand{\pt}[1]{p_{T#1}}
\newcommand{\dimension}[1]{dimension-#1}
\newcommand{\citere}[1]{Ref.\,\cite{#1}}
\newcommand{\citeres}[1]{Refs.\,\cite{#1}}
\newcommand{\code}{\tt}
\newcommand{\sushiversion}{1.6.0}
\newcommand{\sushi}[1]{{\code SusHi#1}}
\newcommand{\sushinew}{\sushi{\_\sushiversion}}
\newcommand{\abbrev}{\scalefont{.9}}

\newcommand{\api}[1]{\frac{\alpha_s#1}{\pi}}
\newcommand{\eqn}[1]{Eq.\,(\ref{#1})}
\newcommand{\fig}[1]{Fig.\,\ref{#1}}
\newcommand{\figs}[1]{Figs.\,\ref{#1}}
\newcommand{\tab}[1]{Tab.\,\ref{#1}}
\newcommand{\sct}[1]{Section~\ref{#1}}
\newcommand{\scts}[1]{Sections~\ref{#1}}
\newcommand{\dd}{{\rm d}}
\newcommand{\deriv}[2]{\frac{\dd #1}{\dd #2}}
\newcommand{\order}[1]{{\cal O}(#1)}
\newcommand{\lhc}{{\abbrev LHC}}
\newcommand{\qcd}{{\abbrev QCD}}
\newcommand{\sm}{{\abbrev SM}}
\newcommand{\thdm}{{\abbrev 2HDM}}
\newcommand{\theory}{{\abbrev TH}}
\newcommand{\mssm}{{\abbrev MSSM}}
\newcommand{\nmssm}{{\abbrev NMSSM}}
\newcommand{\susy}{{\abbrev SUSY}}
\newcommand{\bsm}{{\abbrev BSM}}
\newcommand{\pdf}{{\abbrev PDF}}
\newcommand{\cp}{{\abbrev CP}}
\newcommand{\lo}{{\abbrev LO}}
\newcommand{\nlo}{{\abbrev NLO}}
\newcommand{\nnlo}{{\abbrev NNLO}}
\newcommand{\nklo}[1]{{\abbrev N$^{#1}$LO}}

\newcommand{\msbar}{\overline{\mbox{\abbrev MS}}}

\newcommand{\muF}{\mu_\text{F}}
\newcommand{\muR}{\mu_\text{R}}
\newcommand{\lrr}{l_\text{R0}}
\newcommand{\mphi}{M_\phi}
\newcommand{\mhiggs}{M_\text{H}}
\newcommand{\mtop}{M_\text{t}}
\newcommand{\mbottom}{M_\text{b}}

\newcounter{notecount}
\makeatletter
\def\ps@pprintTitle{%
 \let\@oddhead\@empty
 \let\@evenhead\@empty
 \def\@oddfoot{}%
 \let\@evenfoot\@oddfoot}
\makeatletter
 \def\@oddhead{}
 \let\@evenhead\@empty
 \def\@oddfoot{\centerline{\thepage}}%
 \let\@evenfoot\@oddfoot
\makeatother

\journal{Computer Physics Communications}

\begin{document}
\begin{frontmatter}

\title{
\vspace*{-6em}
  \begin{flushright}
    {\sf\small DESY 16-061, KA-TP-14-2016, TTK-16-14}
  \end{flushright}
\vspace*{2em}
{\sf SusHi Bento}: {\sf Be}yond {\sf N}NLO and the heavy-{\sf to}p limit}

\author{Robert V. Harlander$^1$}
\ead{robert.harlander@cern.ch}

\author{Stefan Liebler$^2$}
\ead{stefan.liebler@desy.de}

\author{Hendrik Mantler$^{3,4}$}
\ead{hendrik.mantler@kit.edu}

\address{
$^1$Institute for Theoretical Particle Physics and Cosmology\\
RWTH Aachen University, 52056 Aachen, Germany\\
$^2$DESY, Notkestra\ss e 85, 22607 Hamburg, Germany\\
$^3$Institute for Theoretical Physics (ITP), Karlsruhe Institute of Technology\\
Engesserstra\ss e 7, 76128 Karlsruhe, Germany\\
$^4$ Institute for Nuclear Physics (IKP), Karlsruhe Institute of Technology\\
Hermann-von-Helmholtz-Platz 1, 76344 Eggenstein-Leopoldshafen, Germany
}

\begin{abstract}
Version {\tt \sushiversion{}} of the code \sushi{} is
presented. Concerning inclusive \cp{}-even Higgs production in
gluon fusion, the following new features with respect to previous
versions have been implemented: expansion of the partonic cross section
in the soft limit, i.e.\ around $x=\mhiggs^2/\hat s\to 1$; \nklo{3}
\qcd{} corrections in terms of the soft expansion; top-quark mass
suppressed terms through \nnlo{}; matching to the cross section at
$x\to 0$ through \nklo{3}. For \cp{}-even and -odd scalars, an
efficient evaluation of the renormalization-scale dependence is
included, and effects of \dimension{5} operators can be studied,
which we demonstrate for the \sm{} Higgs boson and
for a \cp{}-even scalar with a mass of $750$\,GeV.
In addition, as a generalization of the previously available $b\bar b\to H$
cross section, \sushinew{} provides the cross section for charged and
neutral Higgs production in the annihilation of arbitrary heavy quarks.
At fixed order in perturbation theory, \sushi{} thus allows to obtain
Higgs cross-section predictions in different models to the highest
precision known today. For the \sm{} Higgs boson of $\mhiggs=125$\,GeV,
\sushi{} yields $48.28$\,pb for the gluon-fusion cross section at the
\lhc{} at $13$\,TeV. Simultaneously, \sushi{} provides the
renormalization-scale uncertainty of $\pm 1.97$\,pb.
\end{abstract}


\end{frontmatter}

{\bf PROGRAM SUMMARY}\\
\begin{small}
{\em Program title:} \sushi{}.\\
{\em Licensing provisions:} GNU General Public License 3 (GPL).\\
{\em Programming language:} Fortran 77.\\
{\em Reference of previous
  version:}\\ http://dx.doi.org/10.1016/j.cpc.2013.02.006 .\\ {\em Does
  the new version supersede the previous version?:} Yes. The new version
also includes all features of previous versions.\\ {\em Reasons for the
  new version:}\\ Compared to version {\tt 1.0.0} the newest \sushi{}
version {\tt 1.6.0} now supports the 2-Higgs-Doublet-Model (\thdm{}) and
the next-to-minimal supersymmetric standard model (\nmssm{}).  The
effects of \dimension{5} operators in the calculation of the
gluon-fusion cross section can be studied.  It allows to calculate the
Higgs production cross section from the annihilation of heavy quarks and
includes various new features which improve the gluon-fusion
cross-section prediction and the associated uncertainty estimate.  Links
to external codes {\tt 2HDMC}, {\tt MoRe-SusHi} and {\tt
  MadGraph5\_aMC@NLO} can be established.\\ {\em Summary of
  revisions:}\\ Inclusion of \thdm{}, \nmssm{}; Improvements in the
prediction of the gluon-fusion cross section: Top-quark mass terms up to
next-to-next-to leading order, soft expansion and
next-to-next-to-next-to leading order corrections in the heavy top-quark
effective theory, top squark corrections up to next-to-next-to leading
order; \dimension{5} operators; analytic determination of the
renormalization scale dependence. Inclusion of heavy-quark annihilation
cross sections.  Link to {\tt MoRe-SusHi} for the calculation of
resummed transverse-momentum distributions.\\ {\em Nature of
  problem:}\\ Calculation of inclusive and exclusive Higgs production
cross sections in gluon fusion and heavy-quark annihilation in the \sm{}
and extended Higgs sectors through next-to-leading order \qcd{}, including
(next-to-)next-to-next-to-leading order top-(s)quark contributions and
electro\-weak effects.\\ {\em Solution method:} Numerical Monte Carlo
integration.\\ {\em References:} http://sushi.hepforge.org
\end{small}

\section{Introduction}

Since the year 2012, an important task of particle physics is to fully
measure the properties of the Higgs boson with mass $\mhiggs\approx
125$\,GeV discovered at the Large Hadron Collider
(\lhc{})~\cite{Aad:2012tfa,Chatrchyan:2012xdj}.  At the same time, the
search for additional Higgs bosons, which are predicted in many extended
theories, is among the main missions of the \lhc{} experiments. For this
purpose, the knowledge of the corresponding production cross sections
with high precision is of great relevance. The latest efforts in this
direction are regularly summarized in the reports of the ``\lhc{} Higgs
cross section working
group''\cite{Dittmaier:2011ti,Dittmaier:2012vm,Heinemeyer:2013tqa,deFlorian:2016spz}.

In this paper, we describe the new features that have been implemented
in version {\tt \sushiversion{}} of the program
\sushi{}~\cite{Harlander:2012pb,sushiwebpage}.  \sushi{} is a Fortran
code which calculates Higgs-boson production cross sections through
gluon fusion and bottom-quark annihilation in the Standard Model (\sm),
general Two-Higgs-Doublet Models (\thdm), the Minimal Supersymmetric
Standard Model (\mssm) as well as its next-to-minimal extension
(\nmssm), see \citere{Liebler:2015bka}.\footnote{Other codes to obtain
inclusive Higgs-boson cross sections through gluon fusion in the \sm{} and beyond are described in
\citeres{Spira:1995mt,Bagnaschi:2011tu,Anastasiou:2011pi,Anastasiou:2009kn,
Catani:2007vq,Catani:2008me,Ball:2013bra,Bonvini:2014jma,Bonvini:2016frm}.}
Some of these additions to
\sushi{} directly improve the theoretical predictions of the cross
section; others are provided to allow for more sophisticated uncertainty
estimates of these predictions. The new features are the following:
\begin{itemize}
\item \sushi{} now includes the next-to-next-to-next-to-leading order
  (\nklo{3}) terms for the gluon-fusion cross section of a \cp{}-even
  Higgs boson in the heavy-top limit as described in
  \citeres{Anastasiou:2014lda,Anastasiou:2015ema,Anastasiou:2015yha,Anastasiou:2016cez}.
\item It provides the so-called soft expansion of the gluon-fusion cross
  section around the threshold of Higgs-boson production at 
  $x\equiv\mphi^2/\hat s=1$, where $\hat{s}$ denotes the partonic
  center-of-mass energy and $\mphi$ the Higgs-boson mass. This expansion
  is available for the cross sections in the heavy-top limit up to
  \nklo{3} for \cp{}-even Higgs bosons. At next-to-leading order (\nlo{}) and
  next-to-\nlo{} (\nnlo{}), the exact $x$-dependence is still available,
   of course, and remains the default.
\item In addition, \sushinew{} includes top-quark mass effects to the
  gluon-fusion cross section of a \cp{}-even Higgs boson in the
  heavy-top limit up to \nnlo{}, implemented through an expansion in
  inverse powers of the top-quark mass as described in
  \citeres{Marzani:2008az,Harlander:2009my,Harlander:2009mq,%
    Harlander:2009bw,Pak:2009dg,Pak:2009bx,Pak:2011hs}.
  The exact top-mass dependence at lowest order can be factored out.
  We remark that this feature is most interesting at \nnlo{}, of course,
  since at leading order (\lo{}) and \nlo{}, \sushi{} also provides the full quark-mass
  dependence.
\item A matching of the soft expansion to the high-energy
  limit~\cite{Marzani:2008az,Harlander:2009my,Harlander:2009mq},
  i.e.\ $x\to 0$, is available through \nklo{3}.
\item The renormalization-scale dependence of the gluon-fusion cross
  section within an arbitrary interval is calculated in a single
  \sushi{} run.
\item The effect of \dimension{5} operators to the gluon-fusion cross
  section can be taken into account through \nklo{3} \qcd{} for the inclusive
  cross section, and at \lo{} and \nlo{} (i.e.\ $\alpha_s^3$) for the Higgs
  transverse momentum ($\pt{}$) distribution and (pseudo)rapidity
  distribution, respectively.
\item Higgs-boson production cross sections through heavy-quark
  annihilation are implemented along the lines of
  \citere{Harlander:2015xur}, both for the \nnlo{} \qcd{} inclusive
  cross section, as well as for more exclusive
  cross sections up to \nlo{} \qcd{}.
\end{itemize}

All of the described features are applicable to Higgs-boson production
in the theoretical models currently implemented in \sushi{}, even though
some only work for low Higgs masses below the top-quark threshold $\mphi
< 2\mtop$ or for \cp{}-even Higgs bosons.

Our paper is organized as follows: We start with a brief general
overview of the code \sushi{} in \sct{sec:sushi}, and subsequently
present the new features implemented for the prediction of the
gluon-fusion cross section in \sct{sec:gluonfusion}. This includes a
theoretical description of the soft expansion, the inclusion of
\nklo{3} terms and the top-quark mass effects in
\scts{sec:softexp}--\ref{sec:mt}. We proceed with a description of the
``\rge{} procedure'' to determine the renormalization-scale dependence
of the gluon-fusion cross section in \sct{sec:scaledep}, and finally
describe the implementation of an effective Lagrangian including
\dimension{5} operators in \sct{sec:dim5}. The implementation of
heavy-quark annihilation cross sections is described in
\sct{sec:heavyquark}. Numerical results are presented in
\sct{sec:numerics}; they also include a comparison of our results with
the most recent literature. In \ref{app:inputfile} we
present a collection of example input blocks of \sushi{}, which contain
example settings for the various input entries introduced in previous and the newest release.

\section{The program \sushi{}}
\label{sec:sushi}

\sushi{} is a program originally designed to describe Higgs production
in gluon fusion and bottom-quark annihilation in the \mssm{}.  It
collects a number of results from the literature valid through \nklo{3}
in the strong coupling constant, and combines them in a consistent way.
We subsequently discuss the present theoretical knowledge of the
calculation of the gluon fusion and bottom-quark annihilation cross
sections and their inclusion in \sushi{}.

It is well-known that \qcd{} corrections to the gluon-fusion process
$gg\rightarrow \phi$~\cite{Georgi:1977gs}, mediated through heavy quarks
in the \sm{}, are very large. \nlo{} \qcd{} corrections are known for
general quark
masses~\cite{Djouadi:1991tka,Dawson:1990zj,Spira:1995rr,Harlander:2005rq,Anastasiou:2006hc,Aglietti:2006tp}.
In the heavy-top limit, an effective theory can be constructed by
integrating out the top quark. In this case, \nnlo{} corrections have
been calculated a long time
ago~\cite{Harlander:2002wh,Anastasiou:2002yz, Ravindran:2003um}.  The
\nklo{3} contributions were only recently obtained in
\citeres{Anastasiou:2014vaa,Li:2014afw,Anastasiou:2014lda,Anastasiou:2015ema,
  Anastasiou:2016cez}, while various parts of the \nklo{3} calculation
have been calculated
independently\,\cite{Hoschele:2012xc,Baikov:2009bg,Lee:2010cga,Gehrmann:2010ue,
  Gehrmann:2010tu,Gehrmann:2011aa,Kilgore:2013gba,Duhr:2014nda,
  Anastasiou:2015yha,Duhr:2013msa,Anastasiou:2013srw,Hoschele:2014qsa,
  Dulat:2014mda,Anzai:2015wma,Li:2014bfa}.  Approximate \nklo{3} results
were presented in
\citeres{Ball:2013bra,Bonvini:2014jma,deFlorian:2014vta}.  Effects of a
finite top-quark mass at \nnlo{} were approximately taken into account
in~\citeres{Harlander:2009my,Harlander:2009mq,Harlander:2009bw,Pak:2009dg,Pak:2009bx,Pak:2011hs,Marzani:2008az}.

Many of these effects can be taken into account in the latest
version of \sushi{}; this will be discussed in detail in
\sct{sec:gluonfusion}. Electroweak
corrections~\cite{Actis:2008ug,Aglietti:2004nj,Bonciani:2010ms} can be
included as well, either in terms of the full \sm{} electroweak
correction factor, or restricted to the corrections mediated by light
quarks, the latter being a more conservative estimate in certain \bsm{}
scenarios. For completeness, we note that effects beyond fixed order
have been addressed through soft-gluon resummation \cite{Catani:2003zt,
  Moch:2005ky,Idilbi:2005ni,Idilbi:2006dg,Ravindran:2006cg,Ahrens:2008nc,
  Schmidt:2015cea,Bonvini:2016frm}, but those are not included in
\sushi{}.

If requested in the input file, \sushi{} uses the \sm{} results
described above also for the \thdm{}, the \mssm{} or the \nmssm{}
through the proper rescaling of the Yukawa couplings.  In supersymmetric
models, also squarks induce an interaction of the Higgs boson to two
gluons.  In the \mssm{}, the corresponding \nlo{} virtual contributions,
involving squarks, quarks and gluinos, are either known in an expansion
of inverse powers of heavy \susy{}
masses~\cite{Degrassi:2010eu,Degrassi:2011vq,Degrassi:2012vt} or in the
limit of a vanishing Higgs mass, see
\citeres{Harlander:2003bb,Harlander:2004tp,Harlander:2005if,Degrassi:2008zj}.
In this limit, even \nnlo{} corrections of stop-induced contributions
are known, see \citeres{Pak:2010cu,Pak:2012xr}; an approximation of
these effects\,\cite{Harlander:2003kf} is included in \sushi{}, see
\citere{Bagnaschi:2014zla}.  Whereas for the \mssm{} \sushi{} relies on
both expansions, for the \nmssm{} the \nlo{} virtual corrections are
purely based on an expansion in heavy \susy{}
masses~\cite{Liebler:2015bka}.
We note that numerical results for the exact \nlo{} virtual
contributions involving squarks, quarks, and gluinos were presented in
\citeres{Anastasiou:2008rm,Muhlleitner:2010nm}, and analytic results for
the pure squark-induced contributions can be found in
\citeres{Anastasiou:2006hc,Aglietti:2006tp,Muhlleitner:2006wx}.

The associated production of a Higgs boson with bottom quarks,
$pp\rightarrow b\bar b\phi$, is of particular relevance for Higgs
bosons, where the Yukawa coupling to bottom quarks is enhanced. This
happens in models with two Higgs doublets, for example, if $\tan\beta$,
the ratio of the vacuum expectation values of the two neutral Higgs
fields, is large. \sushi{} includes the cross section for this process
in the so-called 5-flavor scheme, i.e.\ for the annihilation process
$b\bar b\to \phi$. The inclusive cross section for this process is
implemented at \nnlo{} \qcd{} \cite{Maltoni:2003pn,Harlander:2003ai}; it
is reweighted by effective Yukawa couplings in the model under
consideration. \sushinew{} now also includes general heavy-quark
annihilation cross sections\,\cite{Harlander:2015xur} at \nnlo{} \qcd{},
which we will describe in \sct{sec:heavyquark}.

For completeness we note that \sushi{} can be linked to {\tt
FeynHiggs}~\cite{Heinemeyer:1998yj,Heinemeyer:1998np,Degrassi:2002fi,Frank:2006yh}
and {\tt 2HDMC}~\cite{Eriksson:2009ws} to obtain consistent sets of
parameters in the \mssm{} or the \thdm{}, respectively.

\sushi{} is controlled via an {\abbrev SLHA}-style \cite{Skands:2003cj}
input file. In the following, we will refer to the entries of a {\tt
  Block "NAME"} and their possible values as
\blockentry{NAME}{ENTRY}{=VALUE}. If more than one value is required, we
write \blockentry{NAME}{ENTRY}{=\{VALUE1,VALUE2,$\dots$\}} or, when
referring only to one specific value,
\blockentry{NAME}{ENTRY,1}{=VALUE1}, etc.
In \ref{app:inputfile} we include input blocks with
example settings for the various new input entries.

\section{Higgs production through gluon fusion}
\label{sec:gluonfusion}

The hadronic cross section for Higgs production in gluon fusion can be
written as
\begin{equation}
\begin{split}
\sigma(pp\to H+X) = \sum_{i,j\in\{q,\bar q,g\}}\tilde{\phi}_i\otimes
\tilde{\phi}_j\otimes\hat\sigma_{ij}\,,
\label{eq:convolution}
\end{split}
\end{equation}
where $\phi_i(x,\muF)=\tilde{\phi}_i(x,\muF)/x$ are parton
densities, $q$ ($\bar q$) denotes the set of all (anti-)quarks ($q=t$
and $\bar q=\bar t$ can be neglected), and $\otimes$ is the convolution
defined as
\begin{equation}
\begin{split}
(f\otimes g)(z) \equiv \int_0^1\dd x_1\int_0^1\dd x_2
  f(x_1)g(x_2)\delta(z-x_1x_2)\,.
\end{split}
\end{equation}
The perturbative expansion of the partonic cross section,
\begin{equation}
\begin{split}
\hat\sigma_{ij,\text{\nklo{n}}}&= 
\sum_{l=0}^n\hat\sigma^{(l)}_{ij}\,,
\label{eq:sigmanij}
\end{split}
\end{equation}
can be represented in terms of Feynman diagrams where the external
partons couple to the Higgs bosons through a top-, bottom-, or
charm-quark loop (contributions from lighter quarks are
negligible).

The first two terms in the perturbative expansion of $\hat\sigma_{ij}$
($l=0,1$ in \eqn{eq:sigmanij}) are known for general quark mass and
included in \sushinew.\footnote{We focus on the \sm{} contributions
  here, but also \susy{} contributions can be added in the first two
  terms and partially even at \nnlo{}, see the description in
  \sct{sec:sushi}.} For the top-quark contribution, the
\nnlo{} term $\hat\sigma^{(2)}_{ij}$ has been evaluated on the basis of
an effective Higgs-gluon interaction vertex which results from
integrating out the top quark from the \sm{} Lagrangian. At \nlo{}, it
has been checked that this results in an excellent approximation of the
\nlo{} \qcd{} correction factor to the \lo{} cross section, even for
rather large Higgs-boson masses.  At \nnlo{}, the validity of the
heavy-top limit for the \qcd{} corrections factor was investigated
through the calculation of a number of terms in an expansion around
$\mtop^2\gg \hat s,\mphi^2$, and matching it to the high-energy limit of
$\hat\sigma^{(2)}_{ij}$\,\cite{Harlander:2009my,Harlander:2009mq,Harlander:2009bw,Pak:2009dg,Pak:2009bx,Pak:2011hs,Marzani:2008az}.
It was found that the mass effects to the \qcd{} correction factor are
at the sub-percent level.

Recently, also the \nklo{3}-term $\hat\sigma_{ij}^{(3)}$ has become
available in terms of a soft expansion and assuming the heavy-top
limit. We will comment on its implementation in the latest release of
\sushi{} in \sct{sec:n3lo}.

The exact \nlo{} and the approximate higher order results for the cross
section are combined in \sushi{} through the formula
\begin{equation}
\begin{split}
\sigma_\text{\X} &= \sigma_\text{\nlo} +
\Delta_\text{\X} \sigma^t\,,\qquad\Delta_\text{\X}
\sigma^t \equiv (1+\delta_\text{EW})\sigma^t_\text{\X} - \sigma_\text{\nlo}^t\,,
\label{eq:sigmasushi}
\end{split}
\end{equation}
where $\sigma_\text{\nlo}$ refers to the \nlo{} cross section with exact
top-, bottom- and charm-mass dependence, while $\sigma^t_\text{\X}$
(\X$=$\nklo{n}, $n\geq 1$) is obtained in the limit of a large top-quark
mass.  Electroweak effects\,\cite{Actis:2008ug}, encoded in
$\delta_\text{EW}$, are included by assuming their full factorization
from the \qcd{} effects, as suggested by \citere{Anastasiou:2008tj} for
a \sm{} Higgs boson.  In \bsm{} scenarios, this assumption may be no
longer justified. \sushi{} therefore provides an alternative way to
include electroweak effects which is based solely on the light-quark
contributions to the electroweak correction factor; for details, we
refer the reader to \citeres{Bagnaschi:2014zla,Harlander:2012pb}. For
our purpose, it suffices to assume \eqn{eq:sigmasushi}. The new release
of \sushi{} provides various approximations to evaluate
$\sigma^t_\text{\X}$, in particular through expansions in $1/\mtop$, and
expansions around $\hat s=\mphi^2$.

In addition to $\sigma_\text{\X}$, which can be found in {\tt Block
  SUSHIggh}, \sushi{} also outputs the individual terms of
\eqn{eq:sigmasushi}. The exact \lo{} and \nlo{} cross sections are
collected in {\tt Block XSGGH}, while the $\sigma^t_\text{\X}$ are given
in {\tt Block XSGGHEFF}, which also contains the electroweak correction
term $\delta_\text{EW}$, if requested.

It is understood that the \nklo{n} terms in \eqn{eq:sigmasushi} are
evaluated with \nklo{n} \pdf{}s.\footnote{Since \nklo{3} \pdf{}s are not
yet available, we use \nnlo{} \pdf{}s for the evaluation of the
\nklo{3} cross section in this paper. The user of \sushi{} can specify
the \pdf{} set at each order individually.} Note that this means that,
for example, $\Delta_\text{\nnlo}\sigma^t$ is not simply the convolution
of $\hat\sigma^{t,(2)}$ with \nnlo{} \pdf{}s, but retains a sensitivity
to $\hat\sigma^{t,(1)}$. Thus, the final result for the \nnlo{}
gluon-fusion cross section obtained from \sushi{} through
\eqn{eq:sigmasushi} depends on the approximation applied to the
evaluation of both $\hat\sigma^{t,(2)}$ and $\hat\sigma^{t,(1)}$. If
electroweak effects are included, this even holds for \sushi{}'s final
result for $\sigma_\text{\nlo}$ due to the definition of
$\Delta_\text{X}\sigma^t$ in \eqn{eq:sigmasushi}.

In the remainder of this section, we first discuss the soft expansion
around the threshold of Higgs production, $\hat{s}=\mphi^2$, in
\sct{sec:softexp}. The implementation of \nklo{3} contributions is
described in \sct{sec:n3lo}, and top-quark mass effects through \nnlo{}
as well as the matching to the high-energy limit in \sct{sec:mt}. While
these features are only available for \cp{}-even Higgs bosons (partially
in a certain range of Higgs-boson masses $\mphi$ only), the analytic
calculation of the $\muR$ dependence of the gluon-fusion cross section
described in \sct{sec:scaledep} is available for all Higgs bosons. The
inclusion of \dimension{5} operators is discussed in \sct{sec:dim5}.

\subsection{Soft expansion}\label{sec:softexp}

The \nlo{} and \nnlo{} coefficients of $\sigma^t$ are approximated very
well by the first few terms\footnote{The first $16$ terms in
 this expansion lead to an accuracy of better than $1$\% with respect to the heavy-top
 limit with exact $x$-dependence at \nnlo{}, for example. For more details see
 below.} in an expansion around the ``soft limit'', $x\to 1$. In fact,
the gain of the full $\hat s$-dependence becomes doubtful anyway when
working in the heavy-top limit, since the latter formally breaks down
for $\hat s>4\mtop^2$, meaning $x\lesssim 0.13$ for $\mhiggs=125$\,GeV.
Apart from the exact $\hat s$-dependence at \lo{}, \nlo{},
and \nnlo{}, \sushinew{} provides the soft expansion of the cross
section for {\abbrev CP}-even Higgs production through order
$(1-x)^{16}$ at these perturbative orders, and also at \nklo{3} (for
more details on the latter, see \sct{sec:n3lo}).

The precise way in which the soft expansion is applied is governed by
the new {\tt Block GGHSOFT}. Each line in this block contains four
integers:
\begin{lstlisting}
Block GGHSOFT
  <entry> <n1> <n2> <n3>
\end{lstlisting}
Following \sct{sec:sushi}, we will refer to such a line as
\blockentry{GGHSOFT}{<entry>}{\tt=\{<n1>,<n2>,<n3>\}} in the text, and
to the individual entries as \blockentry{GGHSOFT}{<entry>,1}{\tt=<n1>},
etc. The integer \blockentry{GGHSOFT}{$n$,1}{} determines
  whether the soft expansion is applied ({\tt =1}) or not ({\tt =0}) at
order \nklo{n}.
  Setting \blockentry{GGHSOFT}{$n$}{=\{1,$\NS,a$\}} evaluates the soft
  expansion of $\hat\sigma^{t,(n)}_{ij}$ in the following way:
\begin{equation}
  \begin{split}
    \hat \sigma^t_{ij}\to
    \hat\sigma^t_{ij,N} \equiv x^{a}{\cal
      T}^x_{\NS}\left(\frac{\Delta\hat\sigma^t_{ij}}{x^{a}}\right)\,,
    \label{eq:softexp}
  \end{split}
\end{equation}
where ${\cal T}^x_\NS$ denotes the asymptotic expansion around $x=1$
through order $(1-x)^\NS$, and $a$ is a non-negative integer. Setting
\blockentry{GGHSOFT}{$n$,2}{=-1} will keep only the soft and collinear
terms, whose $x$ dependence is given by
\begin{equation}
\begin{split}
  \delta(1-x)\quad\text{or}\quad\left(\frac{\ln^k(1-x)}{1-x}\right)_+\,,\quad
  k\geq 0
\end{split}
\end{equation}
by definition.  Here $(\cdot)_+$ denotes the usual plus distribution,
defined by
\begin{equation}
  \begin{split}
    \int_z^1\dd x\left(f(1-x)\right)_+ g(x) = \int_z^1\dd x
    f(x)\left[g(x)-g(1)\right] + g(1)\int_0^z\dd xf(x)\,.
  \end{split}
\end{equation}
The parameters \blockentry{GGHSOFT}{$n$}{} apply to all partonic
subchannels at order \nklo{n}, and to all terms in the $1/\mtop$
expansion as requested by the input {\tt Block GGHMT}, see \sct{sec:mt}
below.

The exact $x$-dependence is obtained by setting
\blockentry{GGHSOFT}{$n$,1}{=0} (only available for $n\leq 2$). The
other two entries in \blockentry{GGHSOFT}{$n$}{} are then irrelevant.
The default values for the block {\tt GGHSOFT} through \nklo{3} are
\begin{equation}
{{\footnotesize
\begin{split}
\fbox{\it default:}\quad
&\text{\blockentry{GGHSOFT}{1,1}{=0}}\,;\quad
\text{\blockentry{GGHSOFT}{2,1}{=0}}\,;\quad
\text{\blockentry{GGHSOFT}{3}{=\{1,16,0\}}}\,.
\end{split}}}
\end{equation}
Again, all terms of the soft expansion are available including the full
$\muF$- and $\muR$-dependence.

A sample input block reads
\begin{lstlisting}
Block GGHSOFT
  1 0 0 0
  2 1 16 1
  3 1 16 1
\end{lstlisting}
which provides the result including the exact $x$-dependence at \nlo{},
and the soft expansion through $(1-x)^{16}$ at \nnlo{} and \nklo{3}
after factoring out a factor of $x$ ($a=1$ in \eqn{eq:softexp}).  We
recall that these settings only affect the heavy-top results
$\hat\sigma^t_\text{\X}$ in \eqn{eq:sigmasushi}; $\sigma_\text{\nlo}$ is
always calculated by taking into account the full quark-mass and
$x$-dependence. The soft expansion is available for all \cp{}-even
Higgs bosons of arbitrary masses.

\subsection{\nklo{3} terms}\label{sec:n3lo}

Recently, the \nklo{3} \qcd{} corrections to the Higgs production cross
section through gluon fusion have become
available\,\cite{Anastasiou:2014lda,Anastasiou:2015ema,Anastasiou:2015yha,Anastasiou:2016cez}. More
specifically, the result was provided in terms of the soft expansion
through order $(1-x)^{37}$ of the leading term in $1/\mtop$ for
$\mu=\muR=\muF$.  We implemented this expansion through $(1-x)^{16}$;
higher order terms do not change the result within the associated
uncertainty. In addition, we included the $\muF$- and $\muR$-dependent
terms at the same order.  Experience from \nnlo{} lets one expect that
these terms are sufficient to obtain an excellent approximation of the
\qcd{} correction factor to the \lo{} cross section, at least for
Higgs masses in the validity range of the effective theory description.

The \nklo{3} result is accessible in \sushinew{} by setting the input
parameter \blockentry{SUSHI}{5}{=3}. This will evolve $\alpha_s(M_Z)$ to
$\alpha_s(\muR)$ at 4-loop order when calculating the cross section,
where $\muR/\mphi$ is defined in \blockentry{SCALES}{1}{}. Note that
with this setting, the hadronic cross section will formally still suffer
from an inconsistency because \nklo{3} \pdf{} sets are not yet
available.
As described in \sct{sec:softexp}, the depth of the soft
expansion at \nklo{3}, as well as the power $a$ in \eqn{eq:softexp} can
be controlled through the input variables
\blockentry{GGHSOFT}{3}{}.\footnote{The setting
  \blockentry{GGHSOFT}{3,1}{=0} is not available, of course.}

Finally, we remark that, also at \nklo{3}, the full $\muR$- and
$\muF$-dependence is available, again accessible through the variables
{\tt SCALES(1)} and {\tt SCALES(2)}, respectively. It follows from
invariance of the hadronic result under these scales, and only requires
the \nnlo{} result as input, as well as the \qcd{} $\beta$ function and
the \qcd{} splitting functions through three loops. The required
convolutions can be evaluated with the help of the program {\tt
  MT.m}\,\cite{Hoeschele:2013gga}, for example.

\subsection{Top-quark mass effects}\label{sec:mt}

In versions before \sushinew{}, only the formally leading terms in
$1/\mtop$ were available for $\hat\sigma^{t,(2)}_{ij}$. However, in
order to allow for thorough studies of the theoretical uncertainty
associated with the gluon-fusion cross section, \sushinew{} includes
also subleading terms in $1/\mtop$ for the production of a \cp{}-even
Higgs (\blockentry{SUSHI}{2}{}$\in\{${\tt
11,12,13}$\}$). There are a number of options provided by
\sushinew{} associated with this; they are controlled by the new input
{\tt Block GGHMT}.

First of all,
\blockentry{GGHMT}{$n$}{=$\NM\in\{$0,1,$\ldots,\NM_n^\text{max}\}$} provides
the expansion of $\hat\sigma^{t,(n)}_{ij}$ through $1/\mtop^{\NM}$ (note
that terms with odd $\NM$ vanish). In addition (or alternatively), one may
define the depth of the expansion individually for each partonic channel
$\hat\sigma^{t,(n)}_{ij}$ through the parameters
\blockentry{GGHMT}{$nm$}{=$\NM\in\{$0,1,$\ldots,\NM_n^\text{max}\}$}, where
$ij=(gg,qg,q\bar q,qq,qq')$ corresponds to $m=(1,2,3,4,5)$,
respectively. Currently, the maximal available depths of expansion
are\footnote{For the $q\bar q$-channel, the maximum reduces to
  $\NM_2^\text{max}=4$, if a soft expansion beyond $\NS=13$ is requested.}
$\NM_0^\text{max}=\NM_1^\text{max}=10$ and $\NM_2^\text{max}=6$.

The default settings are
\begin{equation}
{\footnotesize
\begin{split}
\fbox{\text{\it default:}}\quad 
&\text{\blockentry{GGHMT}{0}{=-1}}\,;\quad
\text{\blockentry{GGHMT}{1}{=0}}\,;\quad
\text{\blockentry{GGHMT}{2}{=0}}\,;\\
&\text{\blockentry{GGHMT}{1$i$}{=GGHMT(1)}}\,,\ i=1,\ldots,3\,;\quad
\text{\blockentry{GGHMT}{2$i$}{=GGHMT(2)}}\,,\ i=1,\ldots,5\,,
\end{split}}
\end{equation}%
where \blockentry{GGHMT}{0}{=-1} means to keep the full top mass
dependence. Let us recall that these settings only affect the heavy-top
results $\hat\sigma^t_\text{\X}$ in \eqn{eq:sigmasushi}; $\sigma_\text{\nlo}$
is always calculated by taking into account the full quark-mass dependence.

As an example, consider the input

\begin{lstlisting}
Block GGHMT
  1    10
  13   0
  2    6
  23   0
  24   0
  25   0
\end{lstlisting}

which will cause \sushinew{} to
\begin{itemize}
\item keep the full top mass dependence at \lo{}
\item expand the \nlo{} terms $\hat\sigma^{t,(1)}_{gg}$ and
$\hat\sigma^{t,(1)}_{qg}$ through $1/\mtop^{10}$
\item expand the \nnlo{} terms $\hat\sigma^{t,(2)}_{gg}$ and
  $\hat\sigma^{t,(2)}_{qg}$ through $1/\mtop^{6}$
\item keep only the terms of order $1/\mtop^0$ for the pure quark
  channels at \nlo{} and \nnlo{}.
\end{itemize}
This also shows that the variables \blockentry{GGHMT}{$nm$}{} overrule
the setting of \blockentry{GGHMT}{$n$}{} for the individual
channels. This may be desirable as it is known that the pure quark
channels show a rather bad convergence
behavior~\cite{Harlander:2009my,Harlander:2009mq,Harlander:2009bw,Pak:2009dg,Pak:2009bx,Pak:2011hs},
so one may want to include only a small number of terms for them in the
$1/\mtop$ expansion.  By convention, \blockentry{GGHMT}{$n$}{} must
always be at least as large as the maximum of \blockentry{GGHMT}{$nm$}{};
if this is not the case in the input file, \sushi{} will override the
user's definition of \blockentry{GGHMT}{$n$}{} and set it to the maximum
of all \blockentry{GGHMT}{$nm$}{}.

In the strict heavy-top limit (i.e., $\NM=0$), the quality of the
approximation improves considerably if one factors out the \lo{} mass
dependence $\sigma^t_0$\,\cite{Graudenz:1992pv,Spira:1995rr} before the
expansion, given by
\vspace{-5mm}
\begin{equation}
\begin{split}
\sigma^t_0 &= \frac{\pi\sqrt{2}G_{\rm F}}{256}\left(\api{}\right)^2
\tau^2\left|1+(1-\tau)\arcsin^2\frac{1}{\sqrt{\tau}}\right|^2\,,\qquad
\tau = \frac{4\mtop^2}{\mphi^2}\,,
\label{eq:sigma0}
\end{split}
\end{equation}
where $G_\text{F}\approx 1.16637\cdot 10^{-5}\,\text{GeV}^{-2}$~\cite{Agashe:2014kda} is
Fermi's constant. The generalization to higher orders in $1/\mtop$
corresponds to
\vspace{-5mm}
\begin{equation}
\begin{split}
\hat\sigma^{t,(n)}_{ij} = \sigma^t_0 \frac{{\cal
    T}_{\NM_{n,ij}}\hat\sigma^{t,(n)}_{ij}}{{\cal T}_{\NM_n}\sigma^t_0}\,,
\label{eq:defexpmt}
\end{split}
\end{equation}
where ${\cal T}_{\NM}$ denotes an operator that performs an asymptotic
expansion through order $1/\mtop^\NM$.  In a strict sense, it should be
$\NM_n=\NM_{n,ij}$; however, \sushi{} allows only for a global value of
$\NM_n$ here, which applies to all sub-channels $ij$ and is set to
\blockentry{GGHMT}{$n$}{}.

Setting \blockentry{GGHMT}{-1}{=$n$} factors out the \lo{} mass
dependence through order $n$, i.e.\
\begin{equation}
\begin{split}
\hat\sigma^t_{ij} &= \sigma^t_0\, \sum_{k=0}^n \frac{{\cal
    T}_{\NM_{k,ij}}\hat\sigma^{t,(k)}_{ij}}{{\cal T}_{\NM_k}\sigma^t_0} +
\sum_{k\geq n+1}{\cal T}_{\NM_{k,ij}}\hat\sigma^{t,(k)}_{ij}\,.
\label{eq:plainexp}
\end{split}
\end{equation}
This will affect all partonic channels. The default setting is
\begin{equation}
{\footnotesize
\begin{split}
\fbox{\text{\it default:}}\quad& \text{\tt GGHMT(-1)=3}
\end{split}}
\end{equation}
which means that the \lo{} $\mtop$ dependence is factored out from
all available orders.

It was observed that higher orders in $1/\mtop$ in general spoil the
validity of the expansion, since its radius of convergence is formally
restricted to $\hat s<4\mtop^2$. This manifests itself in the expansion
coefficients containing positive powers of $\hat s/\mphi^2$. In order to
tame the corresponding divergence as $\hat s\to \infty$, it was
suggested to match the result to the asymptotic behavior in this limit,
which is known from \citeres{Marzani:2008az,Harlander:2009my}. Whether
or not such a matching is performed for $\hat\sigma^{t,(n)}_{ij}$ is
governed by the parameter \blockentry{GGHMT}{$n\cdot$10}{}
(i.e.\ \blockentry{GGHMT}{10}{}, \blockentry{GGHMT}{20}{}, \ldots).  By
default,\\[-2.5em]

\begin{equation}
{\footnotesize%
\begin{split}
\fbox{\text{\it default:}}\quad& \text{\tt
  GGHMT($n\cdot$10)=0}\,,\ n=1,\ldots,3\,, 
\end{split}}
\end{equation}
meaning that no matching is done; setting \blockentry{GGHMT}{$n\cdot$10}{=1}
switches the matching on for all partonic subchannels at order \nklo{n}.

As we will find in \sct{sec:numerics}, the matching to $x\to 0$ is
helpful in approximating the full cross section even at $1/\mtop^0$.
Thus, we provide the possibility to do this matching also at \nklo{3},
even though top-mass suppressed terms are not yet known at this order.
The form of the matching through \nnlo{} has been introduced in
\citeres{Harlander:2009my,Harlander:2009mq}; here we adopt the same
strategy, generalized to \nklo{3}:
\begin{equation}
\begin{split}
\hat\sigma^{t,(n)}_{ij}(x) = \hat\sigma^{t,(n)}_{ij,N}(x)
&+ \,\sigma^t_0 \sum_{l=1}^{n-1}A^{(n,l)}_{ij}\left[ \ln \frac{1}{x} -
  \sum_{k=1}^N\frac{1}{k}(1-x)^k  \right]^l\\&
+ (1-x)^{N+1}\,\left[\sigma^t_0 B_{ij}^{(n)} -
  \hat\sigma^{t,(n)}_{ij,N}(0)
  \right]\,,
\label{eq:match}
\end{split}
\end{equation}
where $\sigma^t_0 B^{(0)}_{ij} = \hat\sigma^{t,(0)}_{ij,N}(0)=0$, and
$\hat\sigma^{t,(n)}_{ij,N}(x)$ denotes the soft expansion of the cross
section through order $(1-x)^N$, see \eqn{eq:softexp}.  The coefficients
$B^{(1)}_{ij}$ and $A^{(2,1)}_{ij}$ are given in numerical form in
\citeres{Marzani:2008az,Marzani:diss,Harlander:2009my},\footnote{The
  notation for $A^{(2,1)}_{ij}$ is $A^{(2)}_{ij}$ in that paper.}  while
$A^{(3,2)}_{gg}$ can be found in \citere{Marzani:diss} (where it is
called $C_\text{A}^3{\cal C}^{(3)}$).  For the unknown coefficients
through \nklo{3}, {\it we assume} 
\begin{equation}
  \begin{split}
\sigma^t_0 B^{(n)}_{ij} &=
\hat\sigma^{t,(n)}_{ij,N}(0)\quad\text{for}\quad n\geq 2\,,\qquad
A^{(3,1)}_{ij}=0\,.
  \end{split}
\end{equation}
The technical consequence of the matching procedure implemented in
\sushi{} is that it {\it requires} the cross section to be expressed in
terms of the soft expansion, i.e., one needs to set
\blockentry{GGHSOFT}{$n$,1}{=1} if the \nklo{n} cross section is requested.

\begin{figure}
\begin{center}
\includegraphics[width=0.47\textwidth]{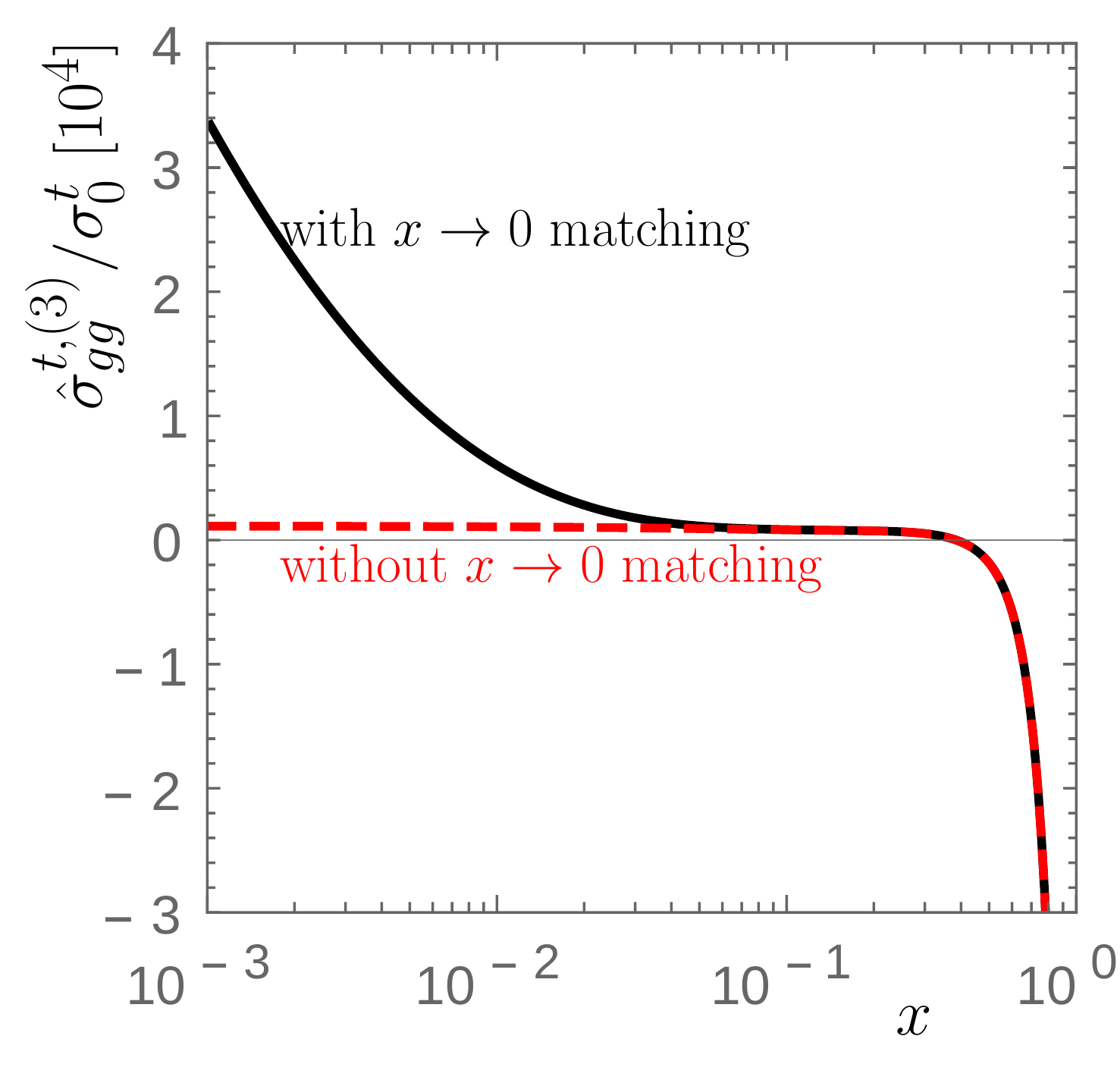}
\end{center}
\vspace{-0.7cm}
\caption{Partonic cross section $\hat{\sigma}_{gg}^{t,(3)}/\sigma^t_0$
  in $10^4$ according to \eqn{eq:match} as a function of
  $x=\mphi^2/\hat{s}$ with and without matching to the high-energy
  limit. The order of the soft expansion applied in both cases is
  $(1-x)^{16}$.}
\label{fig:sgg3match}
\end{figure}

The effect of the matching at \nklo{3} is shown in \fig{fig:sgg3match}:
the soft expansion tends to a constant towards $x\to 0$ by construction,
and cannot reproduce the $\ln^2x$-behavior of the exact result. The
merging of the two limits is very smooth and suggests that the matched
curve is not too far from the full result. Of course, the fact that some
coefficients in \eqn{eq:match} are unknown introduces a theoretical
uncertainty. However, we observe a change in the final cross section of
only about 0.5\% when setting $A^{(3,1)}_{gg}=A^{(3,2)}_{gg}$ for a
\sm{} Higgs, for example.

It remains to say that all terms in $1/\mtop$ are available including
the full $\muF$- and $\muR$-dependence. As in earlier versions of
\sushi{}, these parameters are accessible through the input parameters
{\tt SCALES(1)} and {\tt SCALES(2)}.  Since top-quark mass effects
are not known for the \nklo{3} cross section, all settings
of {\tt GGHMT} involving $n=3$ except from
\blockentry{GGHMT}{30}{} have no effect in the current version
\sushinew{}. The inclusion of $1/\mtop$ terms is only
available for $\mphi<2\mtop$ and the matching to the
high-energy limit only in a mass range $\mphi \in
\left[100\,\text{GeV},300\,\text{GeV}\right]$.

\subsection{Renormalization scale dependence}\label{sec:scaledep}

The renormalization scale ($\muR$) dependence of the partonic cross
section can be written as
\begin{equation}
\begin{split}
\hat\sigma_{ij} = \sum_{n\geq
  0}\sum_{l=0}^n\left(\api{(\muR)}\right)^{n+2}
\hat\kappa^{(n,l)}_{ij}(\mu_0)\,\lrr^l\,,
\label{eq:murdepparton}
\end{split}
\end{equation}
where $\lrr=2\,\ln(\muR/\mu_0)$, and $\mu_0$ is an arbitrary reference
scale. The coefficients $\hat\kappa^{(n,l)}_{ij}(\mu_0)$ are explicitly
contained in \sushi{} (for $\mu_0=\mphi$).  The dependence of the
cross section on $\muR$ can be studied with \sushi{} by varying the
input parameter \blockentry{SCALES}{1}{}, which contains the numerical
value for $\muR/\mphi$. \sushi{} will then insert this value into
\eqn{eq:murdepparton} and convolve the resulting partonic cross section
over the \pdf{}s. A decent picture of the $\muR$ dependence may require
to perform this ``standard procedure'' ten times or more.

\sushinew{} provides a considerably faster way to obtain the $\muR$
dependence of the cross section by convolving the
$\hat\kappa^{(n,l)}_{ij}(\mu_0)$ with the \pdf{}s {\it before} varying
$\muR$,
\begin{equation}
  \begin{split}
    \kappa^{(n,l)}(\mu_0)=
    \hat\kappa^{(n,l)}_{ij}(\mu_0)\otimes \tilde\phi_i\otimes \tilde\phi_j\,.
    \label{eq:kappahad}
  \end{split}
\end{equation}
We will refer to this as the ``{\abbrev RGE} procedure''.  Due to the
renormalization group equation\footnote{The power $n+3$
  takes into account the fact that the \lo{} cross section is of order
  $\alpha_s^2$.}
\begin{equation}
\begin{split}
\deriv{}{\muR^2}\sigma_\text{\nklo{n}} = \order{\alpha_s^{n+3}}
=\deriv{}{\muR^2}\hat\sigma_{ij,\text{\nklo{n}}}\,,
\label{eq:rginv}
\end{split}
\end{equation}
which holds both at the partonic and the hadronic level, it suffices to
calculate the coefficients $\kappa^{(n,l)}(\mu_0)$ for $l=0$ and $n\leq
2$ if the \nklo{3} result is requested. \sushinew{} does this by
initially assuming
$\mu_0/\mphi=\muR/\mphi=$~\blockentry{SCALES}{1}{} in
\eqn{eq:murdepparton}.  All other coefficients are then determined via
the \qcd{} $\beta$ function, defined through
\begin{equation}
\begin{split}
\deriv{}{\muR^2}\alpha_s(\muR) = \alpha_s(\muR)\beta(\alpha_s)\,,
\qquad \beta(\alpha_s) = - \api{}\sum_{n\geq 0}
\left(\api{}\right)^n
\beta_n\,.
\label{eq:betafun}
\end{split}
\end{equation}
Explicitly, one finds
\begin{equation}
\begin{split}
  \kappa^{(1,1)} &= 2\,\beta_0\,\kappa^{(0,0)}\,,\qquad \kappa^{(2,2)} =
  \frac{3}{2}\,\beta_0\,\kappa^{(1,1)}\,,\qquad \kappa^{(2,1)} =
  2\,\beta_1\,\kappa^{(0,0)} +
  3\,\beta_0\,\kappa^{(1,0)}\,,\\ \kappa^{(3,3)} &=
  \frac{4}{3}\beta_0\,\kappa^{(2,2)}\,,\qquad \kappa^{(3,2)} =
  \frac{3}{2}\,\beta_1\,\kappa^{(1,1)} +
  2\,\beta_0\,\kappa^{(2,1)}\,,\\ \kappa^{(3,1)} &=
  2\,\beta_2\,\kappa^{(0,0)} + 3\,\beta_1\,\kappa^{(1,0)} +
  4\,\beta_0\,\kappa^{(2,0)}\,.
\label{eq:murrec}
\end{split}
\end{equation}
Inserting these coefficients into the hadronic analog of
\eqn{eq:murdepparton}, it is possible to obtain the hadronic cross
section at any value of $\muR$ without any further numerical
integration. Since the $\muR$ dependence is typically much larger than
the $\muF$ dependence for gluon fusion, this feature of \sushi{} saves a
significant amount of computing time when aiming for an estimate of the
theoretical uncertainty of the cross section.

Thus, in addition to the usual output file {\tt <outfile>}, running
\sushinew{} with the standard command\\[-2em]

{\footnotesize
\begin{verbatim}
./bin/sushi <infile> <outfile>
\end{verbatim}}

\vspace*{-1em}
will produce an additional file {\tt <outfile>\_murdep} which contains
the gluon-fusion cross section for several values of $\muR$ in the form
\begin{equation}
\begin{array}{rrrrr}
  \muR\text{/GeV}\quad
  & \sigma_\text{\lo}\text{/pb}\quad
  & \sigma_\text{\nlo}\text{/pb}\quad
  & \sigma_\text{\nnlo}\text{/pb}\quad
 & \sigma_\text{\nklo{3}}\text{/pb}
\end{array}
\end{equation}
where all cross sections are evaluated following
\eqn{eq:sigmasushi}, i.e.\ they potentially contain quark-mass
effects, \susy{} corrections, and/or electroweak effects.  The values
of $\muR$ to be scanned over can be set in {\tt <infile>}
through
\begin{lstlisting}
Block SCALES
    1  <mu0mh>
  102  <min0>  <max0>  <N>
\end{lstlisting}
which will cause \sushinew{} to evaluate the cross section at $N+1$
equidistant points for $\log\muR$ between $\log\mu_\text{min}$ and
$\log\mu_\text{max}$, meaning\footnote{{\tt <N>}=$N$, {\tt
    <min0>}=$\mu_\text{min}/\mu_0$, {\tt <max0>}=$\mu_\text{max}/\mu_0$,
  {\tt <mu0mh>}=$\mu_0/\mphi$.}
\begin{equation}
  \begin{split}
    \muR
    &=
    \mu_\text{min}\left(\frac{\mu_\text{max}}{\mu_\text{min}}\right)^{i/N}\,,\qquad
    i\in\{0,1,\ldots,N\}\,.
  \end{split}
  \label{eq:mursampling}
\end{equation}
In addition, \sushinew{} includes a theoretical error estimate on the
inclusive cross section into the standard output file {\tt <outfile>},
given as the maximal and minimal deviation (in pb) within the interval
$\muR\in [\mu_1,\mu_2]$ from the value at $\muR=\mu_0$, using the
sampled values of $\muR$ defined in \eqn{eq:mursampling}, and the cross
sections at the two boundaries $\muR=\mu_1$ and $\mu_2$. The interval is
specified as
\blockentry{SCALES}{101}{=\{$\mu_{1}/\mu_0$,$\mu_2/\mu_0$\}} (recall
that $\mu_0/\mphi=$\blockentry{SCALES}{1}{}); it defaults to
$[\mu_1,\mu_2]=[\mu_0/2,2\mu_0]$.

We remark that this feature works at all perturbative orders through
\nklo{3}, for any settings in the blocks {\tt GGHMT} or {\tt GGHSOFT},
and for any model under consideration. The only restriction is that all
parameters except for the strong coupling constant need to be defined
on-shell. If this is not the case, \sushinew{} will not produce {\tt
  <outfile>\_murdep}. Note that, due to \eqn{eq:sigmasushi}, the
procedure implemented in \sushinew{} is a slightly refined version of
the one described above. In particular, this implies that the \nnlo{}
$\muR$ dependence is {\it exact}, since it is fully determined by the
exact \nlo{} cross section $\sigma_\text{\nlo}$. On the other hand, the
renormalization-scale dependence at \nklo{3} derived from the {\abbrev
  RGE} procedure inherits whatever approximations were made (or not
made) at \nnlo{}.  Thus, the results obtained through the standard and
the {\abbrev RGE} procedure are usually not identical. For example, if
one keeps the full $x$-dependence at \nnlo{}, one also obtains the full
$x$-dependence of the $\muR$-terms at \nklo{3} with the {\abbrev RGE}
procedure, while the standard procedure would only provide them in the
soft expansion.

\subsection{Effective Lagrangian - \dimension{5} operators}\label{sec:dim5}

Let us start from a particular well-defined theory \theory{}; in the
current version of \sushi{}, this could be the \sm{}, a general \thdm{},
the \mssm{}, or the \nmssm{}. We may now include additional gauge
invariant \dimension{5} operators to \theory{} which couple the neutral
Higgs bosons of \theory{} to gluons in the following
way\footnote{\cp{}-even and -odd scalars, which couple
through \dimension{5} operators only, can also be studied, see the description after \eqn{eq:yukfactors}.}:
\begin{equation}
\begin{split}
{\cal L} &= {\cal L}_\text{\theory} +
\sum_{i=1}^{N_1}\frac{\alpha_s}{12\pi v}c_{5,1i}\,H_{1i}G^a_{\mu\nu}G^{a,\mu\nu} +
\sum_{i=1}^{N_2}\frac{\alpha_s}{8\pi v}c_{5,2i}\,H_{2i}G^a_{\mu\nu}\tilde
G^{a,\mu\nu}\,.
\label{eq:leff}
\end{split}
\end{equation}
Here, ${\cal L}_\text{\theory}$ is the Lagrangian of the initial theory
\theory{}, $G^a_{\mu\nu}$ is the gluonic field strength tensor with
color index $a$ and Lorentz indices $\mu$ and $\nu$, and $\tilde
G_{\mu\nu}^a\equiv \varepsilon_{\mu\nu\rho\sigma}G^{a,\rho\sigma}$ is
its dual ($\varepsilon^{0123}=+1$). As usual, $\alpha_s$ is the strong
coupling constant and $v$ the \sm{} Higgs-boson vacuum expectation
value, which we express in terms of Fermi's constant
$v=1/\sqrt{\sqrt{2}G_F}$.  $N_1$ and $N_2$ are the numbers of {\abbrev
  CP}-even and {\abbrev CP}-odd Higgs bosons of the theory,
respectively. The particles themselves are generically denoted by
$H_{1i}$ and $H_{2i}$ (cf.\ also Table\,\ref{tab:htype} below).

The $c_{5,ni}$ denote dimensionless Wilson coefficients which are
understood as perturbative series in $\alpha_s$:
\begin{equation}
 c_{5,ni} = \sum_{k=0}^3 \left(\frac{\alpha_s}{\pi}\right)^k c_{5,ni}^{(k)}\,.
\end{equation}
The normalization is such that $c_{5,ni}^{(0)}=1$ corresponds to the
\lo{} contribution of an infinitely heavy up-type quark $u'$ with
\sm{}-like couplings.\footnote{``\sm-like'' refers to the interaction
  Lagrangian ${\cal L}_\text{int} = -(m_{u'}/v)H_{1i}\bar{u'} u'$ for a
  {\abbrev CP}-even, and ${\cal L}_\text{int} =
  -i(m_{u'}/v)H_{2i}\bar{u'}\gamma_5 u'$ for a {\abbrev CP}-odd Higgs
  boson.} The \nlo{} term for a {\abbrev CP}-even Higgs in this case
would be $c_{5,11}^{(1)}=\tfrac{11}{4}$, etc. In a theory that obeys
naturalness, on the other hand, the order of magnitude of the Wilson
coefficients would be $c_{5,ni} = \order{v/\Lambda}$, where $\Lambda$ is
a scale of physics beyond the \sm{}.

The basic structures for the implementation of the effective Lagrangian
in \eqn{eq:leff} have already been present in earlier versions of
\sushi{}. The reason for this is that the very same operators result
from integrating out the top quark or heavy squarks and gluinos from
${\cal L}_\text{\theory}$. In fact, the \nnlo{} corrections due to top
quarks, as well as the \nlo{} corrections due to top, stop, and gluino
are evaluated on the basis of these \dimension{5}
operators.

Thus, \sushinew{} does not implement any new results; it simply re-uses
previously available {\code functions} and {\code subroutines} in order
to extend the gluon-fusion amplitudes to take into account the effect of
the additional terms in \eqn{eq:leff}.  The numerical values for the
coefficients $c_{5,ni}$ in \eqn{eq:leff} are specified through the newly
introduced {\code Block DIM5}.

\begin{table}
\begin{center}
\begin{tabular}{|c|ccc|}
\hline
{\tt <htype>} & \sm & \thdm/\mssm & \nmssm \\
\hline
11 & $H$ & $h$ & $H_1$\\
12 & $-$ & $H$ & $H_2$\\
13 & $-$ & $-$ & $H_3$\\
21 & $A$ & $A$ & $A_1$\\
22 & $-$ & $-$ & $A_2$\\
\hline
\end{tabular}
\caption[]{\label{tab:htype} Assignment of the \sushi{} input parameter
\blockentry{SUSHI}{2}{=<htype>} to the type of Higgs boson in the
various models. A dash ($-$) means that the assignment is not
meaningful; it will lead to a fatal error in \sushi{}. }
\end{center}
\end{table}

For example, within the \mssm{},
\begin{lstlisting}
Block DIM5
   11   1.00000000E-04    # c5h0
   12   4.00000000E-05    # c5H0
   21  -3.00000000E-07    # c5A0
\end{lstlisting}
corresponds to $c_{5,11}^{(0)} \equiv c_{5,h}^{(0)}=10^{-4}$,
$c_{5,12}^{(0)} \equiv c_{5,H}^{(0)}=4\cdot 10^{-5}$, and
$c_{5,21}^{(0)} \equiv c_{5,A}^{(0)}=-3\cdot 10^{-7}$.  Note that
\sushi{} calculates the cross section of only one particular type of
Higgs boson per run (defined in \blockentry{SUSHI}{2}{}),
see \citere{Harlander:2012pb}. Correspondingly, only the pertinent entry in
{\tt Block DIM5} will have an effect on the result, the other
entries will be ignored. The corrections at higher orders are
specified by setting \blockentry{DIM5}{<k><ni>}{} for
coefficients $c_{5,ni}^{(k)}$ with $k\geq 1$.  At \nlo{} the
contribution of an infinitely heavy up-type quark $u'$ is thus
reproduced by setting \blockentry{DIM5}{11}{=1} and
\blockentry{DIM5}{111}{=2.75}.

The scale dependence of the \dimension{5} Wilson coefficient can be
derived from the non-renormalization of the trace anomaly
term\,\cite{Crewther:1972kn,Chanowitz:1972vd,Chanowitz:1972da,Collins:1976yq},
\begin{equation}
  \begin{split}
    \mu^2\deriv{}{\mu^2} \beta(\alpha_s) G_{\mu\nu}G^{\mu\nu} \equiv 0\,,
  \end{split}
\end{equation}
where $\beta(\alpha_s)$ is given in \eqn{eq:betafun}. Since also
$\alpha_sc_{5,1i}G_{\mu\nu}G^{\mu\nu}$ must be scale invariant, this immediately
leads to~\cite{Spira:1995mt,Brooijmans:2016vro}
\begin{equation}
  \begin{split}
    c_{5,1i}(\muR)=c_{5,1i}(\mu_{\phi})
    \frac{(\beta/\alpha_s)|_{\muR}}{(\beta/\alpha_s)|_{\mu_{\phi}}}\,.
    \label{eq:c5evolve}
  \end{split}
\end{equation}
Perturbatively, we can write this as
\begin{equation}
  c_{5,1i}(\muR)=
  \sum_{n\geq 0}\sum_{l=0}^n\left(\frac{\alpha_s(\muR)}{\pi}\right)^n
  c_{5,1i}^{(n,l)}(\mu_{\phi})l_{R\phi}^l\,,
\label{eq:pertc5} 
\end{equation}
with $l_{R\phi}=2\ln(\muR/\mu_{\phi})$,
\begin{equation}
  c_{5,1i}^{(n,0)} = c_{5,1i}^{(n)}(\mu_{\phi})\,,\quad
  c_{5,1i}^{(n,n)} = 0\qquad \forall\ n
\end{equation}
and, through \nnlo{},
\begin{equation}
\begin{split}
  &c_{5,1i}^{(2,1)}=\beta_0c_{5,1i}^{(1,0)}-\beta_1c_{5,1i}^{(0,0)}\,,\\&
c_{5,1i}^{(3,2)}=\beta_0(\beta_0c_{5,1i}^{(1,0)}-\beta_1c_{5,1i}^{(0,0)})\,,\quad
c_{5,1i}^{(3,1)}=2(\beta_0c_{5,1i}^{(2,0)}-\beta_2c_{5,1i}^{(0,0)})\,.
\end{split}
\end{equation}
Setting \blockentry{DIM5}{0}{=1} makes \sushi{} evolve the Wilson
coefficient perturbatively, i.e.\ according to \eqn{eq:pertc5}; this
is the default. On the other hand, one can also employ
\eqn{eq:c5evolve} for the evolution by setting \blockentry{DIM5}{0}{=2},
similar to the implementation in {\tt HIGLU}~\cite{Spira:1995mt}.
The evolution can also be switched off
(i.e.\ $c_{5,1i}(\muR)=c_{5,1i}(\mu_{\phi})$) by setting \blockentry{DIM5}{0}{=0}.
The {\abbrev RGE} procedure described in the previous
section is only applicable for \blockentry{DIM5}{0}{=1}.
\sushi{} will assume the Wilson coefficient provided in the input {\tt
Block DIM5} to be renormalized at $\mu_{\phi}=\mphi$. The corresponding
values at $\muR$ (where $\muR$ is given in
\blockentry{SCALES}{1}{}) are output in {\tt Block DIM5OUT}.

Moreover, the inclusion of \dimension{5} operators is not compatible with
the inclusion of $1/\mtop$ terms, i.e.\ \sushi{} stops if \blockentry{GGHMT}{1}{$\neq$0} or
\blockentry{GGHMT}{2}{$\neq$0}. The \lo{} dependence including quark-mass
effects must not be factored out, i.e. \sushi{} only accepts
the setting \blockentry{GGHMT}{-1}{=-1}, in order not to reweight
the \dimension{5} operator contributions with top-quark mass effects.

We note that through the {\tt Block FACTORS}, which existed also in
earlier versions, \sushi{} allows to alter the couplings of the Higgs
boson to quarks and squarks. Thus, for example additional factors
$\kappa_t$ and $\kappa_b$ for the Higgs-boson coupling to top and bottom
quarks can be chosen. In case of the \sm{} the corresponding Lagrangian
then takes the following form for the \cp{}-even Higgs boson $H_{11}=H$
\begin{equation}
 \mathcal{L}_\text{\theory} \ni -\kappa_t\sqrt{2}\frac{\mtop}{v}t\bar t H -\kappa_b\sqrt{2}\frac{\mbottom}{v}b\bar b H\,.
 \label{eq:yukfactors}
\end{equation}
It is therefore easily possible to perform an analysis as presented in
\citere{Grojean:2013nya} in \sushi{}, where the dependence of the
gluon-fusion cross section on $\kappa_t$ and $c_{5,1i}$ is discussed.
We will later also focus on this dependence for a very boosted Higgs
taking into account the bottom-quark induced contribution in addition.
Moreover, by setting the couplings to quarks and gauge bosons to
zero through the settings in {\tt Block FACTORS} and \blockentry{SUSHI}{7}{=0}, respectively,
also \cp{}-even or -odd scalars beyond the implemented models can be studied.
We will demonstrate this option by providing inclusive cross sections for a scalar with
a mass of $750$\,GeV at the $13$\,TeV \lhc{} in \sct{sec:numdim5}.

\section{Heavy-quark annihilation}\label{sec:heavyquark}

In this section we shortly comment on the implementation of the total
inclusive \nnlo{} Higgs-production cross sections through heavy-quark
annihilation, $\Q'\bar\Q\to \phi$, as described in
\citere{Harlander:2015xur}. Its activation is through the presence of
the {\tt Block QQH} in the input file, which has the following form:
\begin{lstlisting}
Block QQH
         1      <parton1>
         2      <parton2>
        11      <v*y>
        12      <mu>
\end{lstlisting}
Here, {\tt <parton1>}$\in\{1,\dots,5\}$ denotes the initial-state quark
flavor $\Q'$, and {\tt <parton2>}$\in\{-1,\dots,-5\}$ the initial-state
anti-quark flavor $\bar \Q$. {\tt <v*y>} is the $\qqh$ coupling in the
$\msbar$ scheme at scale {\tt <mu>}$=\mu$/GeV, normalized such that the
\sm{} value of the $q\bar qH$ coupling is {\tt <v*y>}$=m_q(\mu)$/GeV.
For further details regarding the implementation in \sushinew{} and
results we refer to \citere{Harlander:2015xur}.

If the {\tt Block QQH} is provided, \sushi{} will not calculate the
gluon-fusion cross section. The calculation of heavy-quark annihilation
cross sections is also compatible with cuts on the (pseudo)rapidity or
transverse momentum of the Higgs boson up to $\mathcal{O}(\alpha_s^3)$, controlled
through the settings in {\tt Block DISTRIB}.
Also $\pt{}$ distributions (\blockentry{DISTRIB}{1}{=1}) can be requested.
Since all quarks
are assumed massless in this approach, the underlying theory is chirally
symmetric. Therefore the results for a scalar and a pseudo-scalar Higgs
are identical and the setting of \blockentry{SUSHI}{2}{} is irrelevant.
Note also that the collision of an up-type quark with a down-type
anti-quark (or vice versa) implies that $\phi$ carries an electric
charge.  The only model dependence of the $\qqh$ cross section as
calculated by \sushi{} is through the setting of the Yukawa coupling in
\blockentry{QQH}{11}, such that a calculation in the \sm{}-mode is
sufficient (\blockentry{SUSHI}{1}{=0}), unless the Higgs mass should be
obtained from some external code like {\tt FeynHiggs}.

Other parameters of the $\qqh$ calculation are determined by the same
input values as they are used for the $b\bar b\phi$ cross section when no input
{\tt Block QQH} is present. In particular, the perturbative order of
$\qqh$ is controlled through \blockentry{SUSHI}{6}{=$n$}, where
$n=1,2,3$ results in the \lo{}, \nlo{}, or \nnlo{} prediction,
respectively, and the renormalization and factorization scales (relative
to $\mphi$) are defined through \blockentry{SCALES}{11}{} and
\blockentry{SCALES}{12}{}, respectively.

\section{Numerical results}\label{sec:numerics}

This section demonstrates the newly implemented features of \sushinew{}
with the help of exemplary numerical results. We start with a discussion
of the convergence of the soft expansion at individual perturbative
orders up to \nklo{3}, proceed with top-quark mass effects in the
effective field-theory approach, move to the {\abbrev RGE} procedure to
determine the renormalization-scale dependence, before we use these
features to provide a prediction for the cross section of the \sm{}
Higgs boson. Finally, we study the effect of higher dimensional
operators to the transverse momentum~$\pt{}$ of the \sm{} Higgs boson
and provide inclusive cross sections for a \cp{}-even scalar with a
mass of $750$\,GeV. For numerical results concerning heavy-quark annihilation, we refer the
reader to \citere{Harlander:2015xur}.

If not stated otherwise, the setup for the numerical evaluations is as
follows: The \lhc{} center-of-mass energy is set to $\sqrt{s}=13$\,TeV,
and the \sm{} Higgs mass to $\mhiggs=125$\,GeV. We employ {\tt
PDF4LHC15}~\cite{Butterworth:2015oua,Dulat:2015mca,Harland-Lang:2014zoa,
Ball:2014uwa,Gao:2013bia,Carrazza:2015aoa,Carrazza:2015hva}
as parton distribution functions
(\pdf{}), where the {\tt (n)nlo\_mc} Monte Carlo is used by default, and
the {\tt (n)nlo\_100} Hessian sets if noted.  Since \nklo{3} \pdf{} sets
are not available, we use the \nnlo{} set also for the evaluation of the
\nklo{3} terms. Nevertheless, in the \nklo{3} calculation, we evolve
$\alpha_s$ at $4$-loop level; using 3-loop running of $\alpha_s$
instead, the final prediction of the cross section for a \sm{} Higgs
boson changes at the level of $10^{-5}$. The remaining input follows the
recommendation of the \lhc{} Higgs cross section working group,
see~\citere{Denner:2047636}. The on-shell charm-quark mass is set to
$m_c^{\rm OS} = 1.64$\,GeV, which is the upper edge of the range given
in \citere{Denner:2047636}.  The central scale choice for the
renormalization and factorization scale is $\muR=\muF=\mhiggs/2$.

Note that the results of
\scts{sec:numsoftexp}--\ref{sec:bestandrenorm} are obtained for
a \sm{} Higgs boson. However, \sushinew{} allows to take into account
the effects of \nklo{3} contributions in the heavy-top limit and
$1/\mtop$ terms to the \nnlo{} contributions for any \cp{}-even Higgs
boson in the implemented models, as long as the mass of the Higgs boson
under consideration is sufficiently light, i.e.\ below $2\mtop{}$.
Effects of \dimension{5} operators (see \sct{sec:dim5} and
\ref{sec:numdim5}), on the other hand, can be taken into account for any
of the neutral Higgs bosons of the implemented models and
\cp{}-even and -odd scalars, which couple through \dimension{5} operators
only.

\subsection{Soft expansion up to \nklo{3}}
\label{sec:numsoftexp}

In this section, we study the behavior of the expansion around the
``soft limit'', $x\rightarrow 1$, for the gluon-fusion cross section,
see also \sct{sec:softexp}. For the sake of clarity, top-quark
mass effects beyond \lo{} will be neglected in this section, although
the \lo{} cross section including the full top-quark mass dependence is
factored out to all orders (i.e.\ we set \blockentry{GGHMT}{-1}{=3}, see
\sct{sec:mt}).
In order to discuss the convergence of the soft expansion, we define the
quantity
\begin{equation}
 \left(\frac{\delta \sigma}{\sigma}\right)^{\text{\nklo{n}}} =
 \frac{\sigma^t_{\text{\nklo{n}},\NS,a}}{\sigma^t_{\text{\nklo{n-1}}}}-1\quad\text{
   with } n\geq 1 \,,
\label{eq:delsigbysig}
\end{equation}
where $\sigma^t_\text{\nklo{n}}$ has been introduced in
\eqn{eq:softexp}.  Through $\order{\alpha_s^{n+1}}$, the exact
$x$-dependence is taken into account. In the highest-order terms,
i.e.\ the terms of order $\order{\alpha_s^{n+2}}$ in
$\sigma^t_{\text{\nklo{n}},\NS,a}$, the soft expansion is applied
according to \eqn{eq:softexp} up to order $(1-x)^\NS$ with $N\leq
16$. All studies in this subsection were performed without matching the
cross section to the result at $x\to 0$, i.e., we set
\blockentry{GGHMT}{$n\cdot$10}{=0} for $n=1,2,3$.

At infinite order of the soft expansion, the value of the parameter $a$
in \eqn{eq:softexp} is obviously irrelevant.  If only a finite number of
terms in the expansion is available, the dependence of the result on the
parameter $a$ has been studied in detail in
\citere{Anastasiou:2016cez}. It was shown that the soft expansion seems
to converge particularly well for small, non-negative values of $a$. The
differences among the final results for different values of $a$ are smaller
at higher orders, as we demonstrate subsequently. One
observes that the $\muF$-dependent terms of $\hat\sigma$ at \nlo{} are
polynomial in $x$, which means that they are {\it identical} to their
soft expansion for $a=0$ once it is taken to sufficiently high order ($N=3$, to be
specific). This is no longer true with the choice $a > 0$. Let us add
that, since the $\muR$-dependent terms at \nlo{} are proportional to
$\delta(1-x)$, they are the same whether the soft expansion is applied
or not.

\begin{figure}
\begin{center}
\begin{tabular}{cc}
\includegraphics[width=0.47\textwidth]{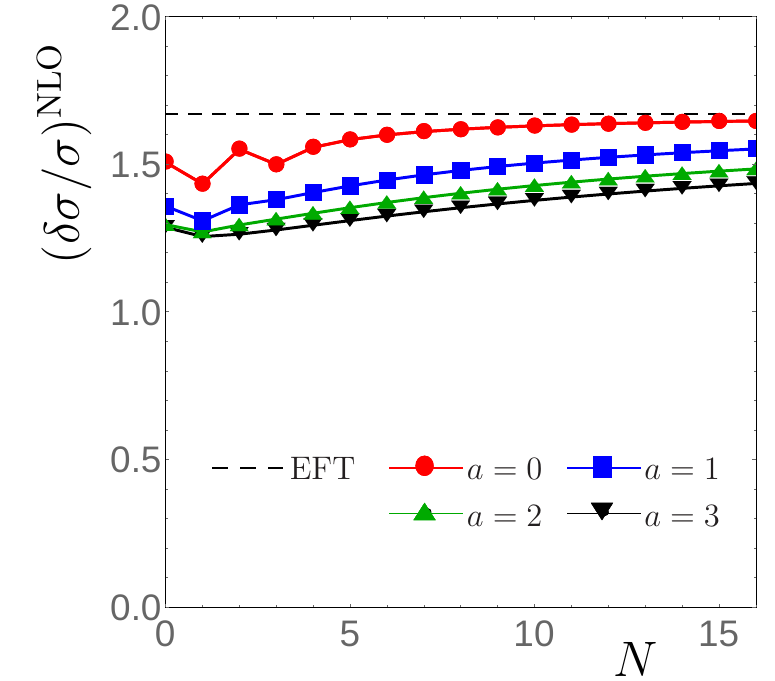} &
\includegraphics[width=0.47\textwidth]{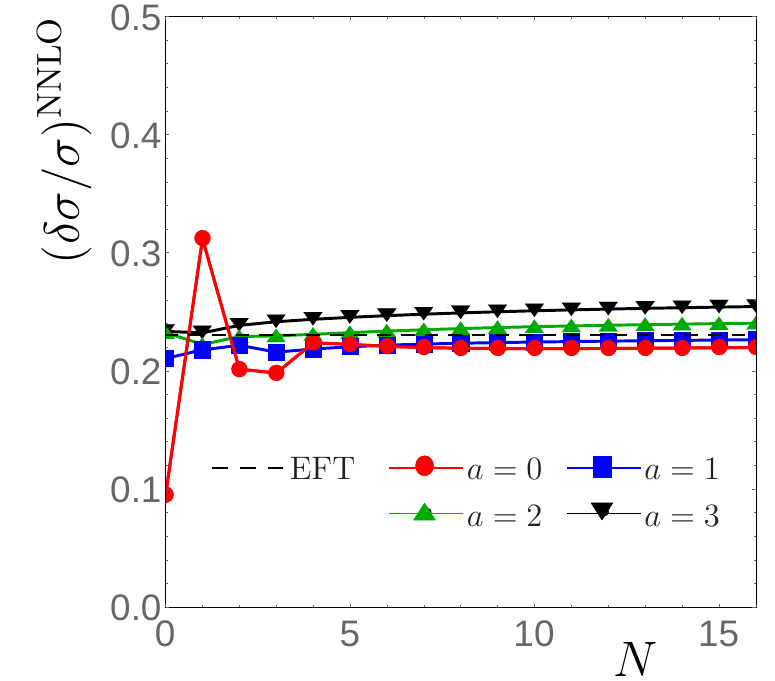}  \\[-0.4cm]
 (a) & (b)
\end{tabular}
\end{center}
\vspace{-0.7cm}
\caption{ (a) Convergence of the \nlo{} cross section as a function of
$\NS$ for $a=0,1,2,3$ in \eqn{eq:softexp}; (b) Convergence of the
\nnlo{} cross section as a function of $\NS$ for $a=0,1,2,3$ in
\eqn{eq:softexp}.  In both figures the colors depict $a=0$ (red),
$a=1$ (blue), $a=2$ (green), $a=3$ (black). The black, dashed line
corresponds to the exact result in the heavy-top limit. 
The results are obtained for a \sm{} Higgs
with $\mhiggs=125$\,GeV at the $\sqrt{s}=13$\,TeV \lhc{}.}
\label{fig:softexp}
\end{figure}

\begin{figure}
\begin{center}
\begin{tabular}{cc}
\includegraphics[width=0.47\textwidth]{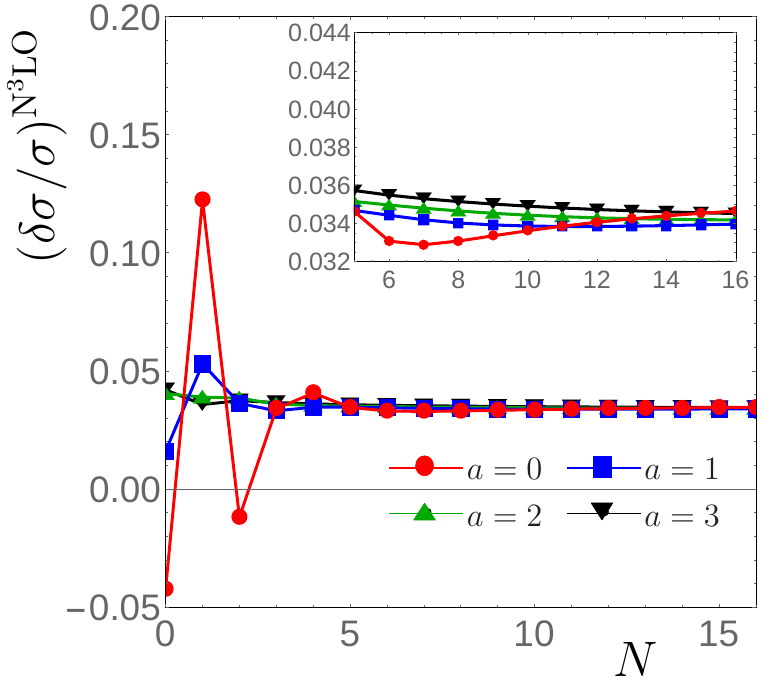} &
\includegraphics[width=0.47\textwidth]{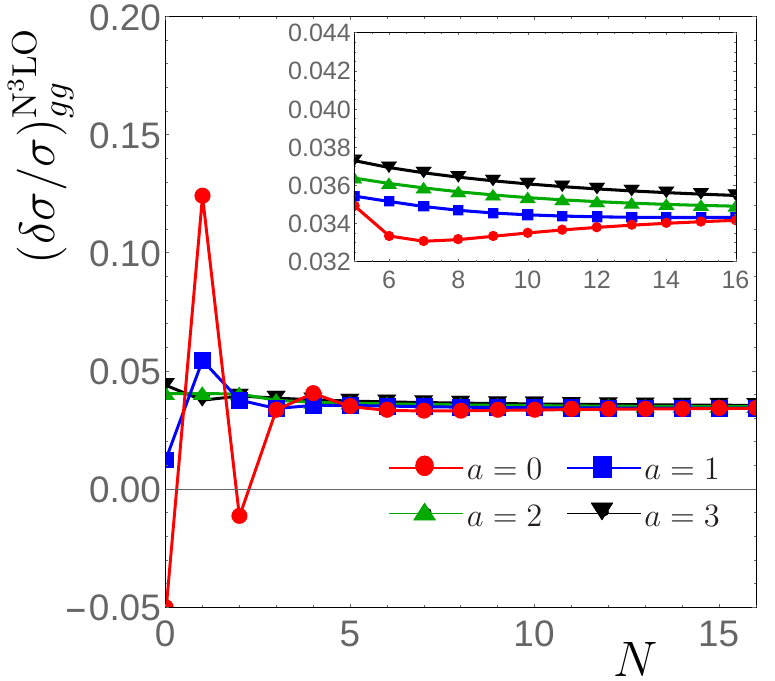}  \\[-0.4cm]
 (a) & (b)
\end{tabular}
\end{center}
\vspace{-0.7cm}
\caption{ (a) Convergence of the \nklo{3} cross section as a
function of $\NS$ for $a=0,1,2,3$ in \eqn{eq:softexp}; (b) Convergence
of the gg channel of the \nklo{3} as a function of $\NS$ for
$a=0,1,2,3$ in \eqn{eq:softexp}.  A zoom for larger values of $\NS$ is
provided in the upper right corner of the figures.  In both figures
the colors depict $a=0$ (red), $a=1$ (blue), $a=2$ (green), $a=3$
(black). The results are obtained for a \sm{} Higgs
with $\mhiggs=125$\,GeV at the $\sqrt{s}=13$\,TeV \lhc{}. }
\label{fig:softexp2}
\end{figure}

\figs{fig:softexp}\,(a) and (b) show the convergence of the soft
expansion at \nlo{} and \nnlo{}, respectively.  Both figures also
include the result without soft expansion as dashed black line,
i.e.\ where $\sigma^t_{\text{\nklo{n}},\NS,a}$ is replaced by
$\sigma^t_{\text{\nklo{n}}}$ in \eqn{eq:delsigbysig}. At \nlo{}, the
case $a=0$ appears to be clearly preferable; for larger value of $a$,
the soft expansion is further away from the exact
$x$-dependence. For $\NS\geq 9$, the deviation for $a=0$ is less
than $2.5$\%.\footnote{Note that this refers to the absolute \nlo{}
{\it correction} term in pb; with respect to the total cross
section, this translates into an approximation which is better than
$1.6$\%.} It decreases down to $1.3$\% at $\NS=16$, while the result
for $a=1$ is still more than $7$\% off.  

At \nnlo, convergence of the soft expansion appears to be a bit faster,
with no significant impact of the terms higher than $(1-x)^6$ both for
$a=0$ and $a=1$. For $\NS\geq 9$, the result for $a=0$ ($a=1$)
approximates the exact $x$-dependence of the correction term to better
than $5$\% ($2$\%) (translating into about $0.9$\% ($0.3$\%) for the total cross
section).

\fig{fig:softexp2}~(a) depicts the convergence of the soft expansion for
the cross section at \nklo{3}. Above $\NS=11$, the spread among the
curves for $a=0,1,2,3$ is of the order of $3$\% of
$\delta\sigma/\sigma$, which means about $0.1$\% of the total cross
section. For completeness, the same plot for the dominant $gg$ channel
alone is shown in \fig{fig:softexp2}~(b). Note that in this case, we
only include the $gg$ channel also in the denominator of
\eqn{eq:delsigbysig}. At lower orders of the soft expansion, the curve
for $a=0$ behaves less smoothly compared to $a\geq 1$; at sufficiently
high orders though, all results can be considered consistent with each
other at the level of accuracy indicated above.

\subsection{Top-quark mass effects through \nnlo{} and matching to the high-energy limit}
\label{sec:topquarkmass}

In this section we comment on top-quark mass effects beyond the
heavy-top limit, which can be taken into account in \sushi{} up to
$1/\mtop^{10}$ at \lo{} and \nlo{} and up to $1/\mtop^6$ at \nnlo{}. As
already pointed out in \sct{sec:mt}, a naive expansion of the
partonic cross section in $1/\mtop$ breaks down. Thus, in this section,
we apply the matching to the high-energy limit as described in
\sct{sec:mt}, i.e.\ we set
\blockentry{GGHMT}{$n\cdot$10}{=1} for $n=1,2,3$.

\begin{figure}
\begin{center}
\begin{tabular}{cc}
\includegraphics[width=0.47\textwidth]{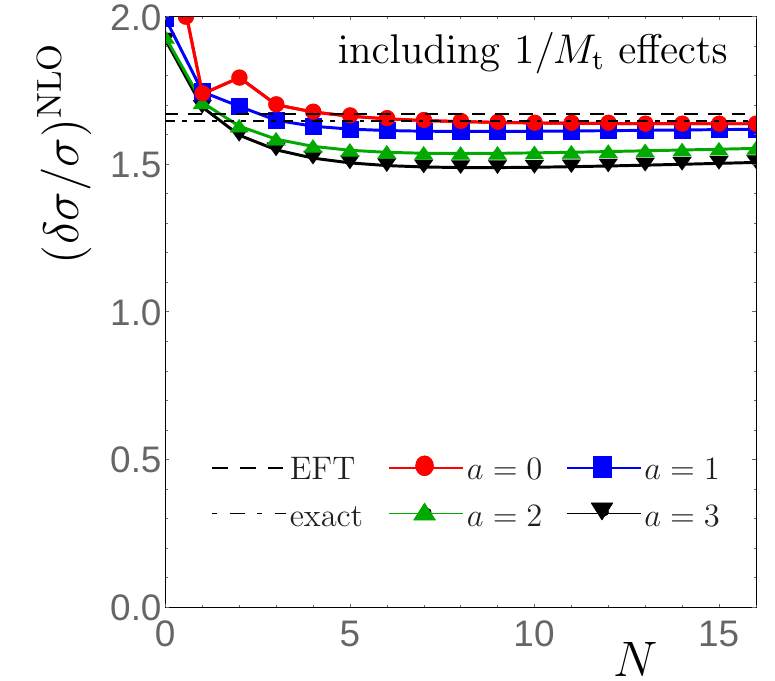} &
\includegraphics[width=0.47\textwidth]{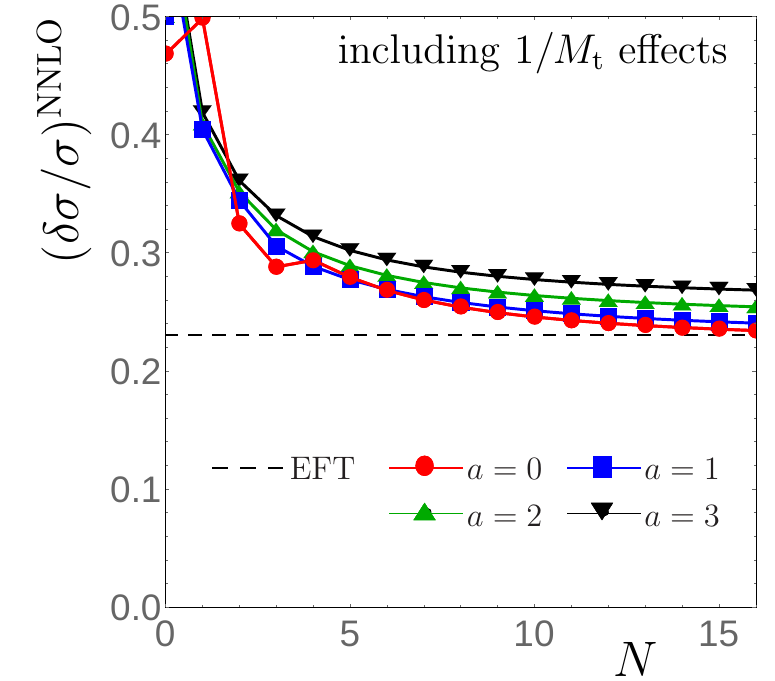}  \\[-0.4cm]
 (a) & (b)
\end{tabular}
\end{center}
\vspace{-0.7cm}
\caption{
(a) Convergence of the \nlo{} cross section as a function of $\NS$ for
$a=0,1,2,3$ in \eqn{eq:softexp} with top-quark mass effects up to $1/\mtop^8$;
(b) Convergence of the \nnlo{} cross section as a function of $\NS$ for
$a=0,1,2,3$ in \eqn{eq:softexp} with top-quark mass effects up to $1/\mtop^4$.
In both figures the colors depict $a=0$ (red), $a=1$ (blue), $a=2$ (green),
$a=3$ (black). The black, dashed line corresponds to
the exact result in the heavy-top limit, the black, dot-dashed line
to the exact result with full top-quark mass dependence (only known at \nlo{}).
The results are obtained for a \sm{} Higgs with $\mhiggs=125$\,GeV
at the $\sqrt{s}=13$\,TeV \lhc{}.
}
\label{fig:softexpmt}
\end{figure}

\begin{figure}
\begin{center}
\begin{tabular}{cc}
\includegraphics[width=0.47\textwidth]{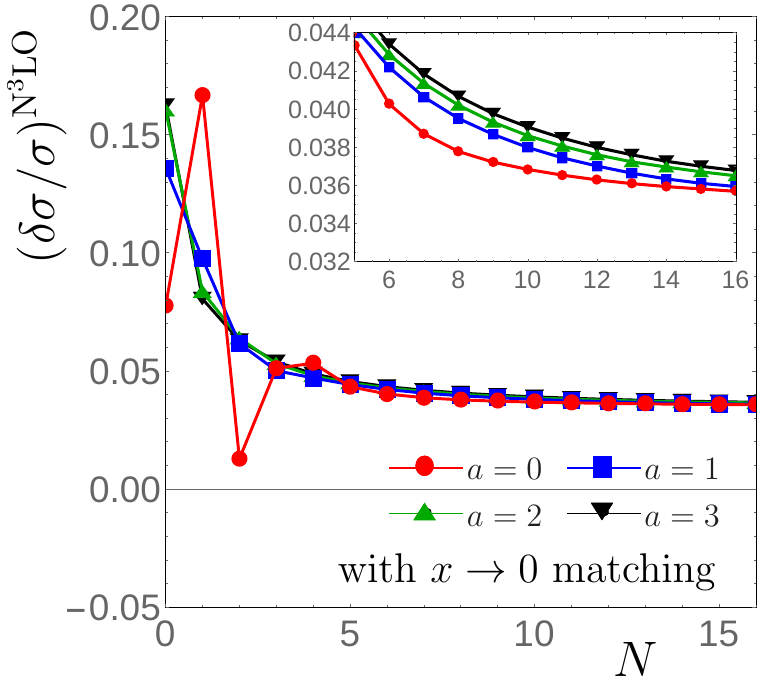} &
\includegraphics[width=0.47\textwidth]{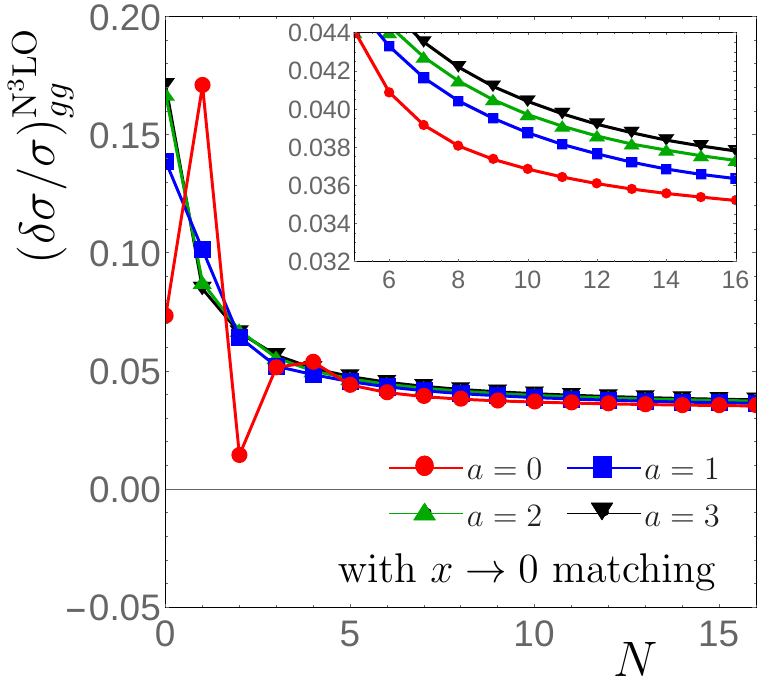}  \\[-0.4cm]
 (a) & (b)
\end{tabular}
\end{center}
\vspace{-0.7cm}
\caption{ (a) Convergence of the \nklo{3} cross section as a
function of $\NS$ for $a=0,1,2,3$ in \eqn{eq:softexp}; (b) Convergence
of the $gg$ channel of the \nklo{3} as a function of $\NS$ for
$a=0,1,2,3$ in \eqn{eq:softexp}.  A zoom for larger values of $\NS$ is
provided in the upper right corner of the figures.  In both figures
the colors depict $a=0$ (red), $a=1$ (blue), $a=2$ (green), $a=3$
(black). In contrast to \fig{fig:softexp2} the \nklo{3} result
is matched to the high-energy limit. 
The results are obtained for a \sm{} Higgs
with $\mhiggs=125$\,GeV at the $\sqrt{s}=13$\,TeV \lhc{}.}
\label{fig:softexpmt2}
\end{figure}

Recall that the matching procedure of
\citeres{Harlander:2009mq,Harlander:2009my} requires the soft expansion
of the partonic cross section. Thus before discussing the relevance of
the top-quark mass effects, it is necessary to study the convergence of
the soft expansion also for these terms.  For the result at \nlo{} we
can compare to the result in the heavy-top limit, but also to the exact
top-quark mass dependence; the difference between these two results is
about $1$\%.  At \nnlo{}, on the other hand, only a comparison to the
heavy-top limit is possible.  The results are shown in
\fig{fig:softexpmt}, including terms through $1/\mtop^8$ at \nlo{}, and
through $1/\mtop^4$ at \nnlo{} (for the $gg$ and the $qg$ channels
also $1/\mtop^6$ terms are implemented in \sushi{} but provide a
negligible contribution, see \fig{fig:mtterms} below).
Following \eqn{eq:delsigbysig}, we keep the exact
$x$-dependence one order below to allow for a better comparison with
the figures of \sct{sec:numsoftexp}. At \nlo{}, one observes a nice convergence of
the soft expansion to the exact result, provided $a=0$. Terms beyond
$(1-x)^{10}$ have only negligible effects on the final result in this
case. At \nnlo{}, convergence of the soft expansion is significantly
slower, but the available number of terms in this expansion seems
sufficient for a prediction of the mass effects with permille level
accuracy, provided that $a=0$ is indeed the most reliable choice for the
parameter defined in \eqn{eq:softexp}.  \fig{fig:softexpmt2} shows
the \nklo{3} result with matching to the high-energy limit as
described in \sct{sec:mt}. The convergence of the soft expansion as a
function of $N$ is slightly worse compared to the result without
matching, but shows a similar behavior as the results at \nlo{} and
\nnlo{} depicted in \fig{fig:softexpmt}. The correction
at $N=16$ is comparable to the result without matching, see
\fig{fig:softexp2}.

Let us now discuss the top-quark mass effects at different orders
$1/\mtop^\NM$ in more detail, while applying the soft expansion through
$(1-x)^{16}$ with $a=0$ (see \eqn{eq:softexp}).  The result is presented
in \fig{fig:mtterms}, where the relative difference
\begin{equation}
  \begin{split}
    \left(\frac{\delta\sigma}{\sigma}\right)_{\mtop}
    =\frac{\sigma_{\NM}}%
    {\sigma_\text{htl}}
    -1
    \label{eq:delsigmt}
  \end{split}
\end{equation}
to the heavy-top limit at the corresponding perturbative order is shown.
At \nlo{}, $\sigma_{\NM}$ is obtained by including terms of order
$1/\mtop^P$ in the partonic cross section and matching it to the $x\to0$
limit (i.e.\ \blockentry{GGHMT}{1}{=$P$}, \blockentry{GGHMT}{10}{=1},
\blockentry{GGHSOFT}{1}{=\{1,16,0\}}), while $\sigma_\text{htl}$ is the
heavy-top limit at \nlo{}
(i.e.\ \blockentry{GGHMT}{1}{=}\blockentry{GGHMT}{10}{=0},
\blockentry{GGHSOFT}{1}{=\{0,0,0\}}). In both cases, the value for the
cross section provided by \sushi{} in \blockentry{XSGGHEFF}{1}{} is used.

At \nnlo{}, we use \eqn{eq:sigmasushi} which corresponds to the \sushi{}
output \blockentry{SUSHIggh}{1}, neglecting bottom- and charm-quark, and
electroweak effects
(\blockentry{FACTORS}{1}{=}\blockentry{FACTORS}{3}{=}\blockentry{SUSHI}{7}{=0}).
Furthermore, we make sure that only the genuine \nnlo{} effects of the
$1/\mtop$ terms are shown, by fixing the approximation used at
$\order{\alpha_s^3}$; specifically, we set \blockentry{GGHMT}{1}{=6},
\blockentry{GGHMT}{10}{=1}, and \blockentry{GGHSOFT}{1}{=\{1,16,0\}}, both for
$\sigma_P$ and $\sigma_\text{htl}$. For the $\mathcal{O}(\alpha_s^4)$-terms, we apply
the analogous settings of the \nlo{} case described above. I.e., we
include terms of order $1/\mtop^P$ in $\sigma_P$ (modulo the restriction
to $1/\mtop^4$ for the pure quark channels, see above), and match them
to the $x\to 0$ limit, while we apply the usual heavy-top limit
for $\sigma_\text{htl}$.

The results are shown in \fig{fig:mtterms}, together with the relative
difference of the {\it exact} \nlo{} cross section to its heavy-top
limit (black dashed).
The points at $\NM=0$ illustrate the effect of using the soft expansion
combined with matching to the result at $x=0$, as opposed to keeping the
full $x$ dependence (without matching). Both at \nlo{} and \nnlo{}, this
effect is obviously larger than the genuine $1/\mtop$-terms.  This
underlines that, as long as one works in a heavy-top approximation,
which is strictly valid only for $x>\mphi^2/(4\mtop^2)$, the full
$x$-dependence is not necessarily an improvement w.r.t.\ the soft
expansion, in particular if additional information like the $x\to 0$
limit is available.

Both at \nlo{} and \nnlo{}, the $1/\mtop$ terms exhibit
a nice convergence behavior. However, the observation at \nlo{} is that,
while the $1/\mtop^0$ result almost exactly reproduces the full mass
dependence after matching to the high-energy limit and employing the soft expansion,
including higher-order mass effects moves the approximation
{\it away} from the exact result. Thus, we cannot expect that their
inclusion at \nnlo{} leads to an improved result w.r.t.\ the heavy-top
limit.  Nevertheless, we believe that their overall behavior allows to
derive an upper bound on the top-mass effects to the heavy-top limit of
the order of
$1$\%\,\cite{Harlander:2009my,Harlander:2009mq,Pak:2009dg}.

\begin{figure}
\begin{center}
\includegraphics[width=0.47\textwidth]{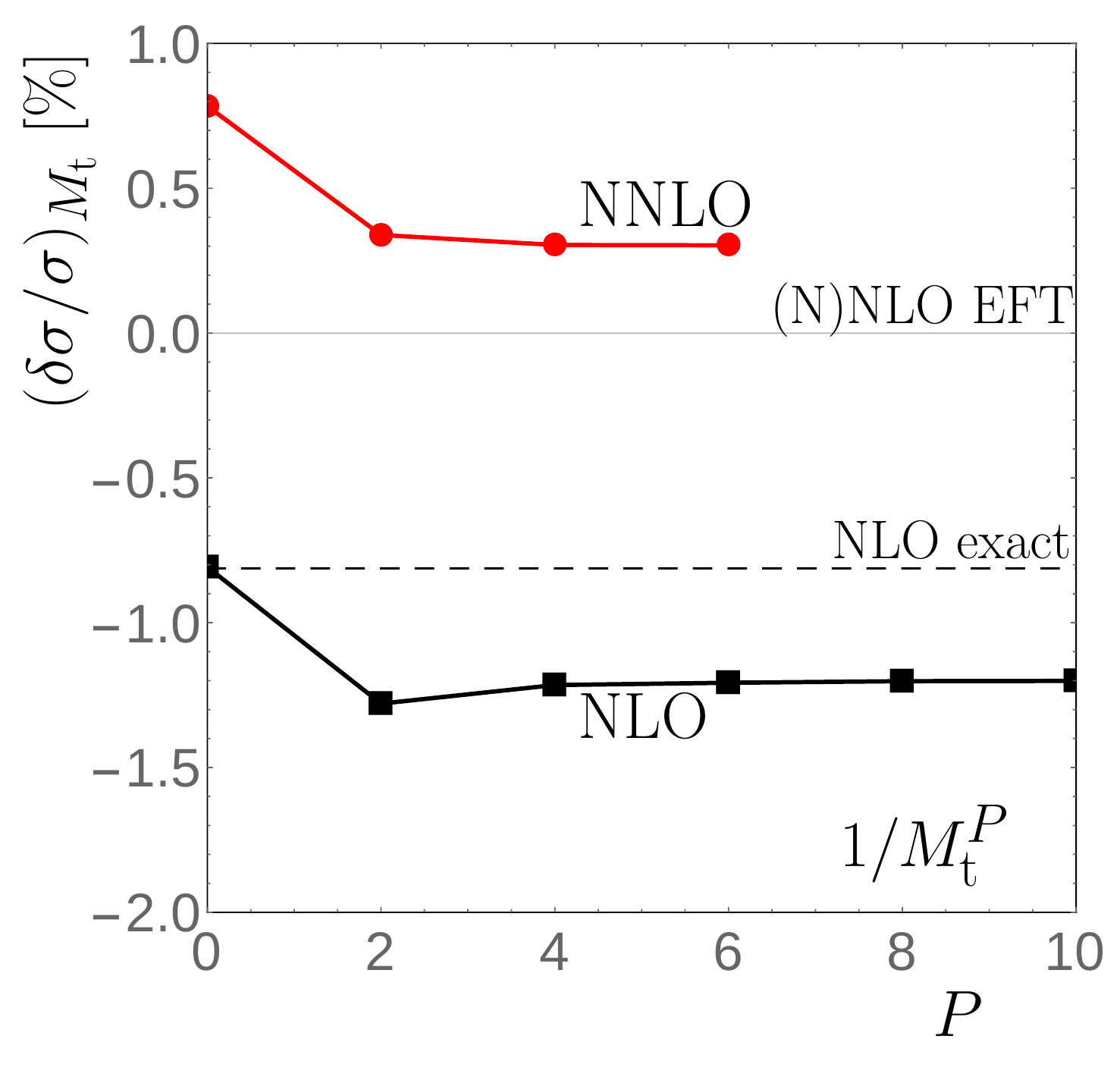}
\end{center}
\vspace{-0.7cm}
\caption{ Relevance of $1/\mtop^\NM$ terms to the cross section in the
heavy-top limit in percent at \nlo{} (black) up to $\NM=10$ and at \nnlo{}
(red) up to $\NM=6$ with respect to the exact heavy-top limit at the
corresponding order.  The black, dashed line corresponds to the exact
\nlo{} result. The results are obtained for a \sm{} Higgs
with $\mhiggs=125$\,GeV at the $\sqrt{s}=13$\,TeV \lhc{}.  }
\label{fig:mtterms}
\end{figure}

\subsection{Cross section prediction for the \sm{} Higgs boson and scale dependence}
\label{sec:bestandrenorm}

Having discussed the top-quark mass terms to the \nlo{} and \nnlo{}
cross section in the heavy-top limit and the convergence of the soft
expansion, we can finally provide a prediction for the cross section of
the \sm{} Higgs boson including its scale uncertainty. In this section
we make use of the Hessian \pdf{} sets {\tt PDF4LHC15\_(n)nlo\_100}.
Following the arguments of the preceding sections, the best prediction
of \sushi{} is obtained with the following settings: use the
perturbative result through \nklo{3}, i.e.\ set
\blockentry{SUSHI}{5}{=3}; at each order of the effective-theory result,
apply the soft expansion through $(1-x)^{16}$ with $a=0$, i.e.\ set
\blockentry{GGHSOFT}{$n$}{=\{1,16,0\}} for $n\in\{1,2,3\}$; take into
account top-quark mass terms to the predictions of the \nlo{} and
\nnlo{} cross sections in the heavy-top limit through the settings
\blockentry{GGHMT}{$n$}{=4} for $n\in\{1,2\}$, i.e.\ $1/\mtop^4$ terms
are taken into account at \nlo{} and \nnlo{};
match to the high-energy limit $x\rightarrow 0$ at \nlo{}, \nnlo{},
and \nklo{3}, i.e.\ set \blockentry{GGHMT}{$n\cdot$10}{=1} for
$n=1,2,3$. The choice of $a=0$ is motivated through the reproduction
of the correct scale dependence at \nlo{} and the observations in
\sct{sec:numerics}. Also note that for all predictions in the effective
field-theory approach, we factor out the full top-quark mass dependence,
i.e.\ \blockentry{GGHMT}{-1}{=3}.  Finally, we include the electroweak
correction factor according to \eqn{eq:sigmasushi}, i.e.\ we set
\blockentry{SUSHI}{7}{=2}.  The exact \nlo{} cross section of
\eqn{eq:sigmasushi} contains the contributions from the three heaviest
quarks: top, bottom, and charm.  The numbers can be reproduced with the
input file {\tt SM-N3LO\_best.in} in the {\tt example}-folder of the
\sushinew{} distribution.

\newpage
With this setup, we obtain
\vspace{-7mm}
\begin{equation}
\begin{split}
 \text{\nnlo{}}:                 &\qquad \sigma = 43.55\,\text{pb}\pm 4.44\,\text{pb}(\muR)\,,\\
 +\text{\nklo{3}}:               &\qquad \sigma = 45.20\,\text{pb}\pm 1.61\,\text{pb}(\muR)\,,\\
 +\text{$1/\mtop$ effects at \nlo{} and \nnlo{}}\,\, &\\
 +\text{matching ($x\to 0$) at \nlo{}, \nnlo{} and \nklo{3}}:&\qquad \sigma = 45.80\,\text{pb}\pm 1.87\,\text{pb}(\muR)\,,\\
 +\text{electroweak corrections}:&\qquad \sigma = 48.28\,\text{pb}\pm 1.97\,\text{pb}(\muR)\,, 
 \label{eq:res}
\end{split}
\end{equation}
where the uncertainty $\pm \Delta(\muR)$ only takes into account the
renormalization-scale dependence. Here, $\Delta(\muR)$ is the maximum
deviation of the cross section within the interval $\muR/\mhiggs
\in[1/4,1]$ from the value at $\muR=\mhiggs/2$. Each line of
\eqn{eq:res}, including the uncertainty, has been obtained in a single
run of \sushi{}, which takes a few seconds on a modern desktop
computer. The final result is perfectly consistent within its
uncertainties with the prediction $48.58\,\text{pb}\pm 1$\,pb($\muR$) given
in \citere{Anastasiou:2016cez} and the result $48.1\,\text{pb}\pm 2.0$\,pb
(without resummation) employing the Cacciari-Houdeau Bayesian approach~\cite{Cacciari:2011ze}
to estimate higher unknown orders presented in \citere{Bonvini:2016frm}.
We note that the result of
\citere{Anastasiou:2016cez} was computed with the \nnlo{} \pdf{} set at
all orders, whereas we employ the \nlo{} \pdf{} set for the \nlo{} terms
in \eqn{eq:sigmasushi}.  If we employ {\tt PDF4LHC15\_nnlo\_100} instead
at all orders, we obtain $48.37$\,pb.  Other uncertainties need to be
added as described in \citeres{deFlorian:2016spz,Anastasiou:2016cez}.

\begin{figure}
\begin{center}
\begin{tabular}{cc}
\includegraphics[width=0.47\textwidth]{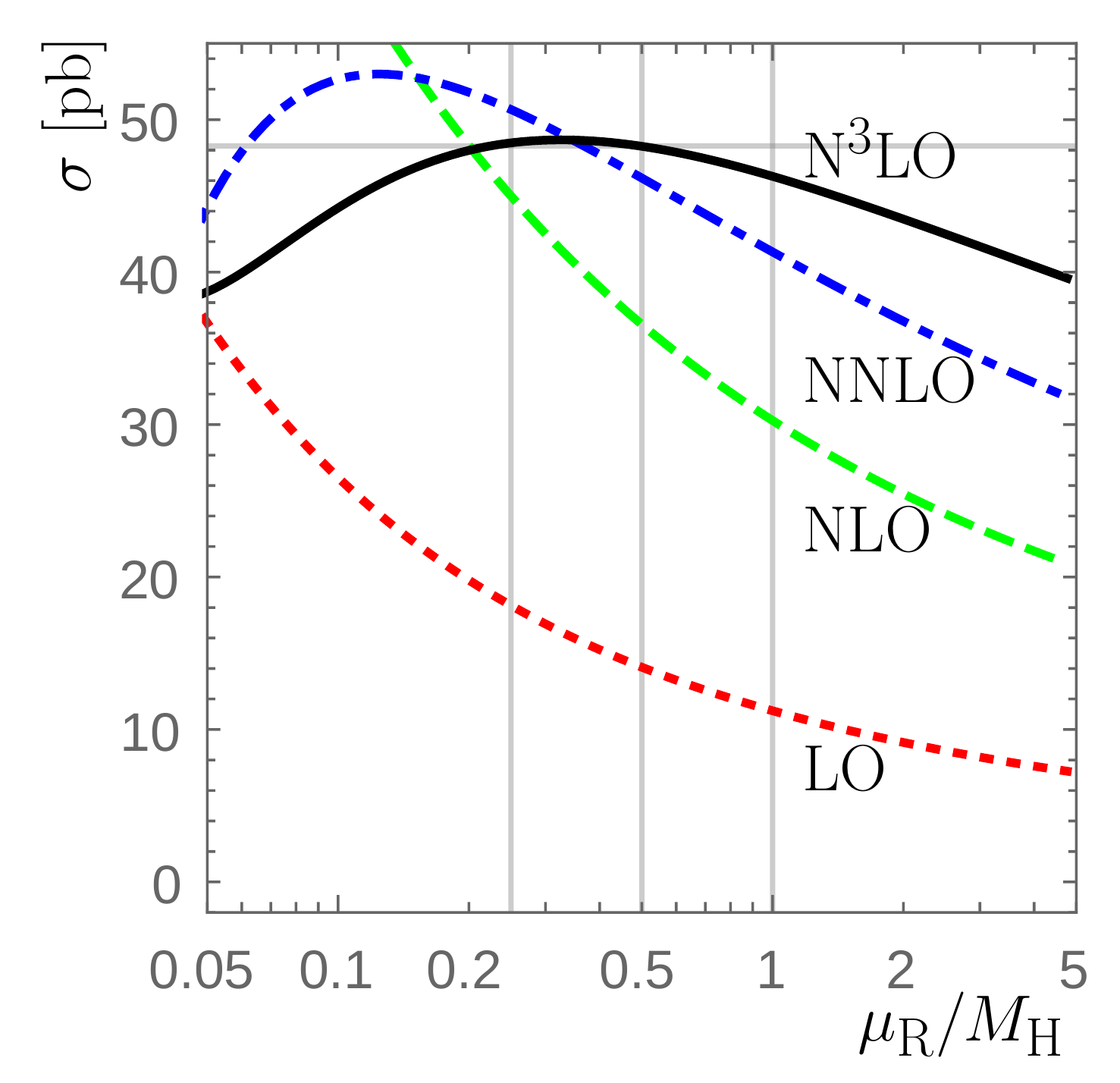} &
\includegraphics[width=0.47\textwidth]{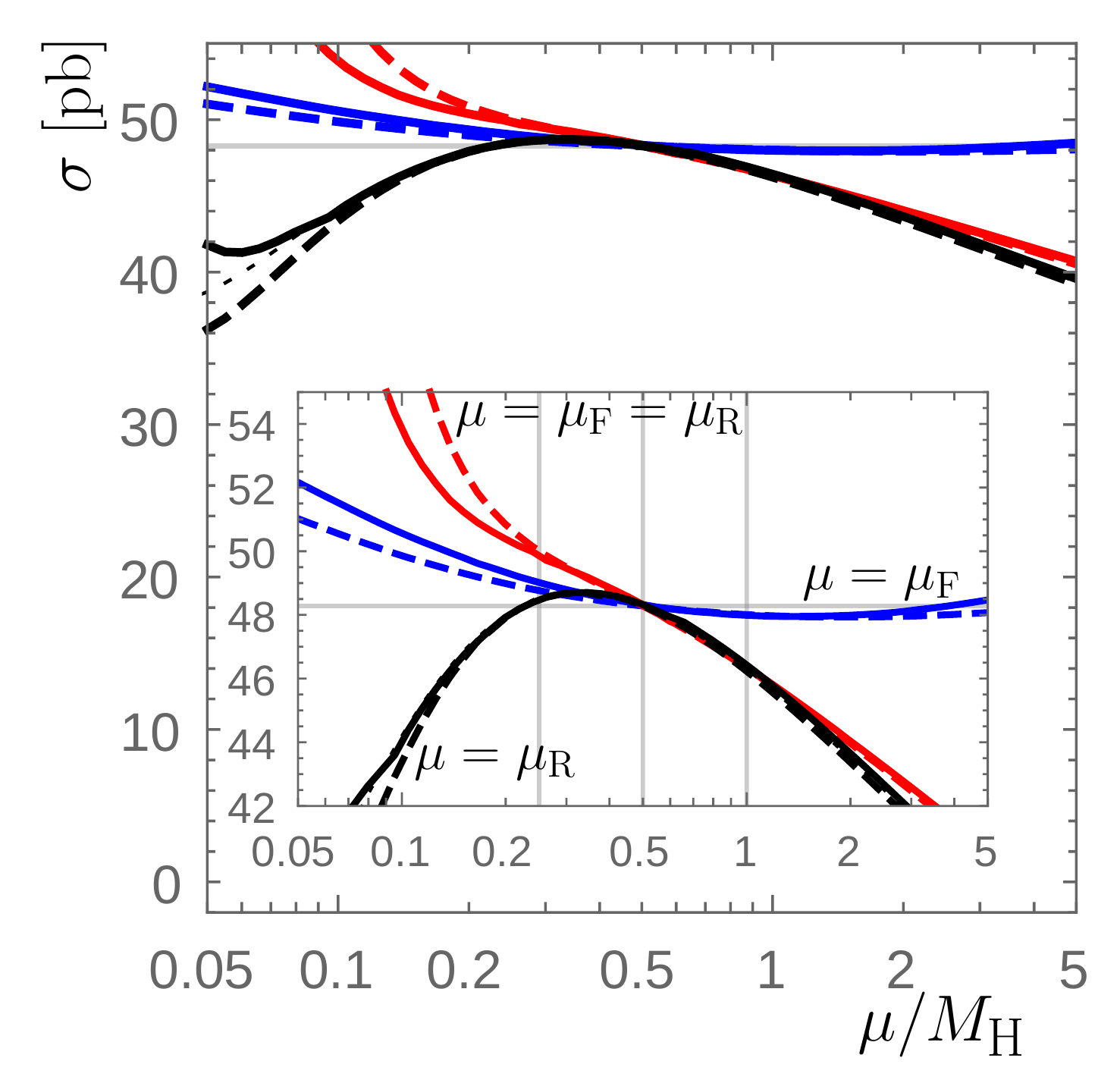}  \\[-0.4cm]
 (a) & (b)
\end{tabular}
\end{center}
\vspace{-0.7cm}
\caption{ (a) \lo{} (red, dotted), \nlo{} (green, dashed), \nnlo{}
(blue, dot-dashed) and \nklo{3} (black, solid) gluon-fusion cross
section in pb (see \eqn{eq:sigmasushi}) as a function of
$\muR/\mhiggs$ (obtained in a single run); (b) Best prediction cross
section in pb as a function of $\muF/\mhiggs$ (together with
$\muR=\mhiggs/2$) (blue) and $\muF/\mhiggs=\muR/\mhiggs$ (red) and
$\muR/\mhiggs$ (together with $\muF=\mhiggs/2$) (black). Each curve
is shown twice, once for $a=0$ (solid) and $a=1$ (dashed) in the soft
expansion at \nklo{3}. The dotted, thin black line depicts the \nklo{3}
result from (a). Both figures are obtained for a \sm{} Higgs
with $\mhiggs=125$\,GeV at the $\sqrt{s}=13$\,TeV \lhc{}. }
\label{fig:scaledep}
\end{figure}

Running the input file {\tt SM-N3LO\_best.in} also generates a file
including the renormalization-scale dependence. Its content is shown in
\fig{fig:scaledep}~(a). The dependence clearly
reduces successively from \nlo{} to \nklo{3}.  Note that at
each order we follow \eqn{eq:sigmasushi} and thus include the
electroweak correction factor beyond \lo{}.  The flat behavior around
$\muR=\mhiggs/2$ leads to a highly asymmetric scale variation around the
central value, suggesting a symmetrization of the corresponding
uncertainty band as done in \eqn{eq:res}.  As explained in
\sct{sec:scaledep}, the $\muR$ dependence obtained through the {\abbrev
  RGE} procedure at \nklo{n} is as precise as the calculation at
\nklo{n-1}, while in the standard procedure (by manually varying
\blockentry{SCALES}{1}{}), its precision is determined by the \nklo{n}
calculation.  We show the result of the standard procedure in
\fig{fig:scaledep}~(b) (black lines). In addition, the $\mu=\muF$
dependence for $\muR=\mhiggs/2$ (blue) and the combined $\mu=\muF=\muR$
dependence (red) are shown. In each case, the solid and dashed line
corresponds to setting $a=0$ and $a=1$ in \eqn{eq:softexp},
respectively.  The differences between these two cases, as well as
between the standard and the {\abbrev RGE} procedure are small, except for small
values of $\mu$. We also observe that the behavior at low values of $\mu$ in
\fig{fig:scaledep}~(b) is dependent on the soft expansion and the
matching performed at \nlo{} and \nnlo{}.
However, within the interval $\mu\in\left[\mhiggs/4,\mhiggs \right]$
which we use for the uncertainty determination, the agreement is good.

\subsection{Dimension~$5$ operators}
\label{sec:numdim5}

\begin{figure}[htp]
\begin{center}
\begin{tabular}{cc}
\includegraphics[width=0.47\textwidth]{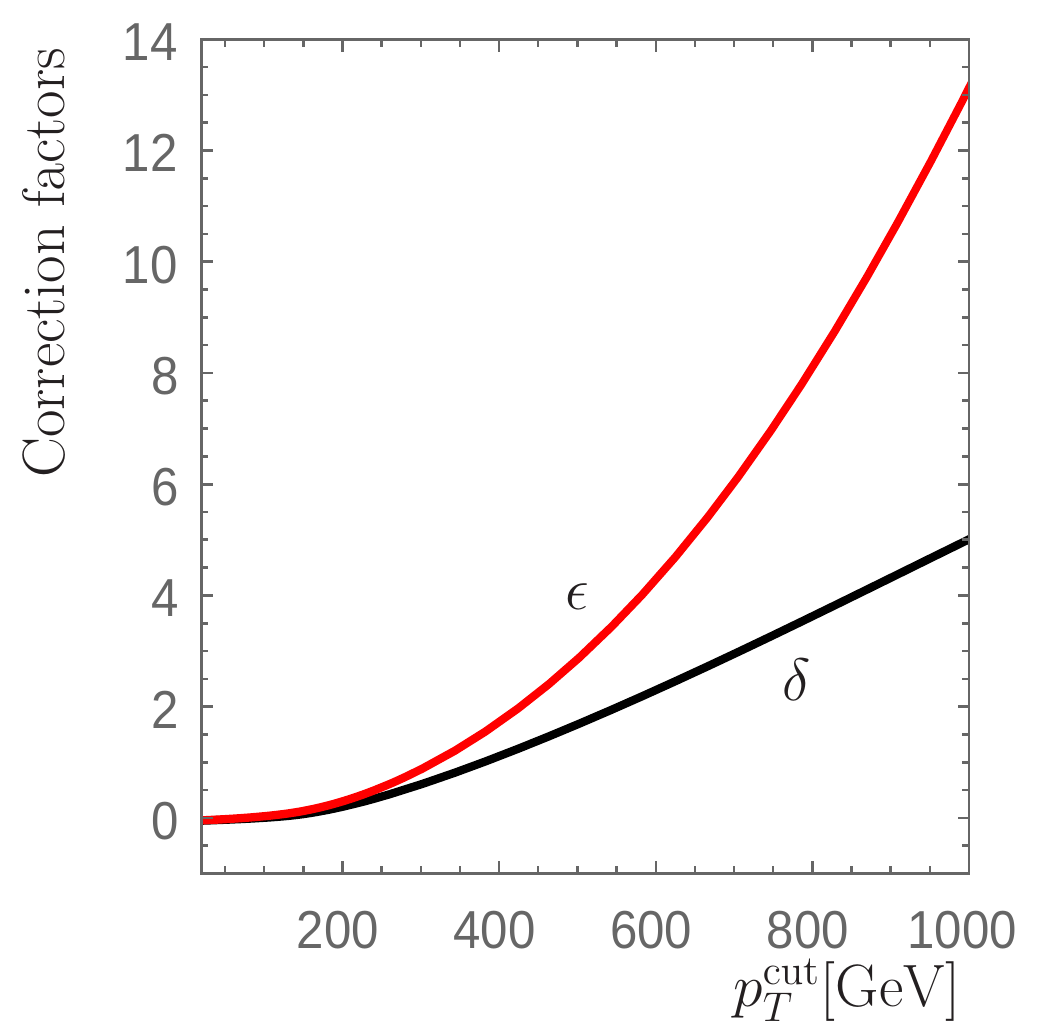} &
\includegraphics[width=0.47\textwidth]{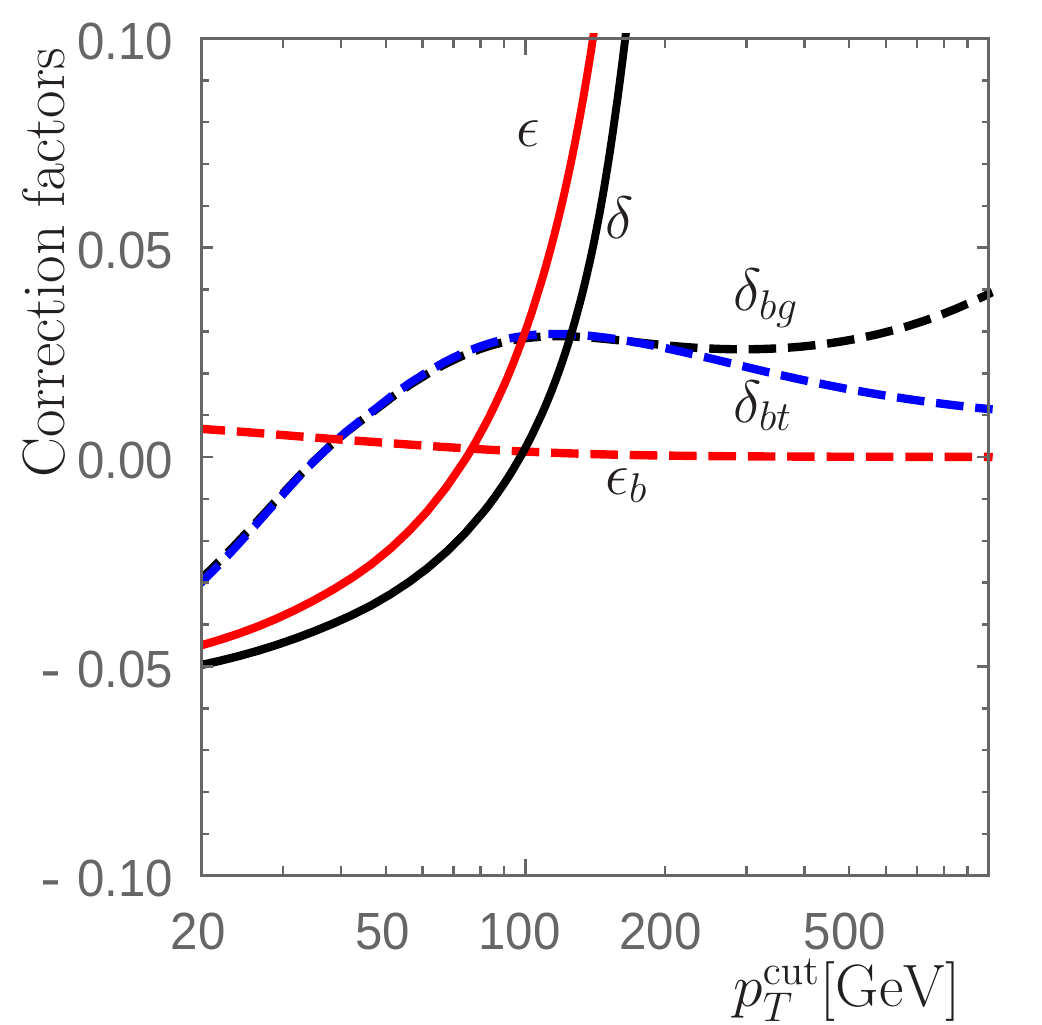}  \\[-0.4cm]
 (a)  & (b)
\end{tabular}
\end{center}
\vspace{-0.7cm}
\caption{(a) Correction factors $\delta$, $\epsilon$ as a function of the
lower cut $\pt{}^\text{cut}$ in GeV and in addition (b) $\delta_{bg}$, $\delta_{bt}$ and $\epsilon_{b}$
as a function of $\pt{}^\text{cut}$. Both figures are obtained for a \sm{}
Higgs with $\mhiggs=125$\,GeV at the $\sqrt{s}=13$\,TeV \lhc{}.}
\label{fig:corrfactors} 
\end{figure}

In order to study the effect of the \dimension{5} operators, it is
helpful to consider the fraction of events where the \sm{} Higgs boson is
produced at transverse momenta above a certain value~$\pt{}^\text{cut}$. We define
\begin{equation}
\begin{split}
R(\pt{}^\text{cut})
=\frac{1}{\sigma^\text{tot}}\sigma(\pt{}^\text{cut})\quad\text{with}\quad
\sigma(\pt{}^\text{cut})
=\int_{\pt{}>\pt{}^\text{cut}}
\dd\pt{}\frac{\dd\sigma}{\dd\pt{}}\,,
\label{eq:ptrat}
\end{split}
\end{equation}
where $\sigma\equiv \sigma_{ni}(c_{5,ni})$ denotes the cross section for
the production of a Higgs boson $H_{ni}$ within the theory defined by
\eqn{eq:leff}, and follow the numerical setup described at the beginning
of \sct{sec:numerics}.  However, we do not take into account charm-quark
and electroweak contributions and choose a $\pt{}$-dependent
renormalization and factorization scale for the result presented in
\fig{fig:corrfactors}.
If not stated otherwise, the relative Yukawa couplings to top- and
bottom quarks are set to one, i.e.\ we discuss the specific model
\theory{} with additional \dimension{5} operator.
In the subsequent \nlo{} analysis, we set $c_{5}^{(1)}=\tfrac{11}{4}c_{5}^{(0)}$,
i.e.\ our \dimension{5} operator assumes the same (rescaled) \nlo{}
correction as for the top-quark induced Wilson coefficient.

\begin{figure}
\begin{center}
\begin{tabular}{cc}
\includegraphics[width=0.47\textwidth]{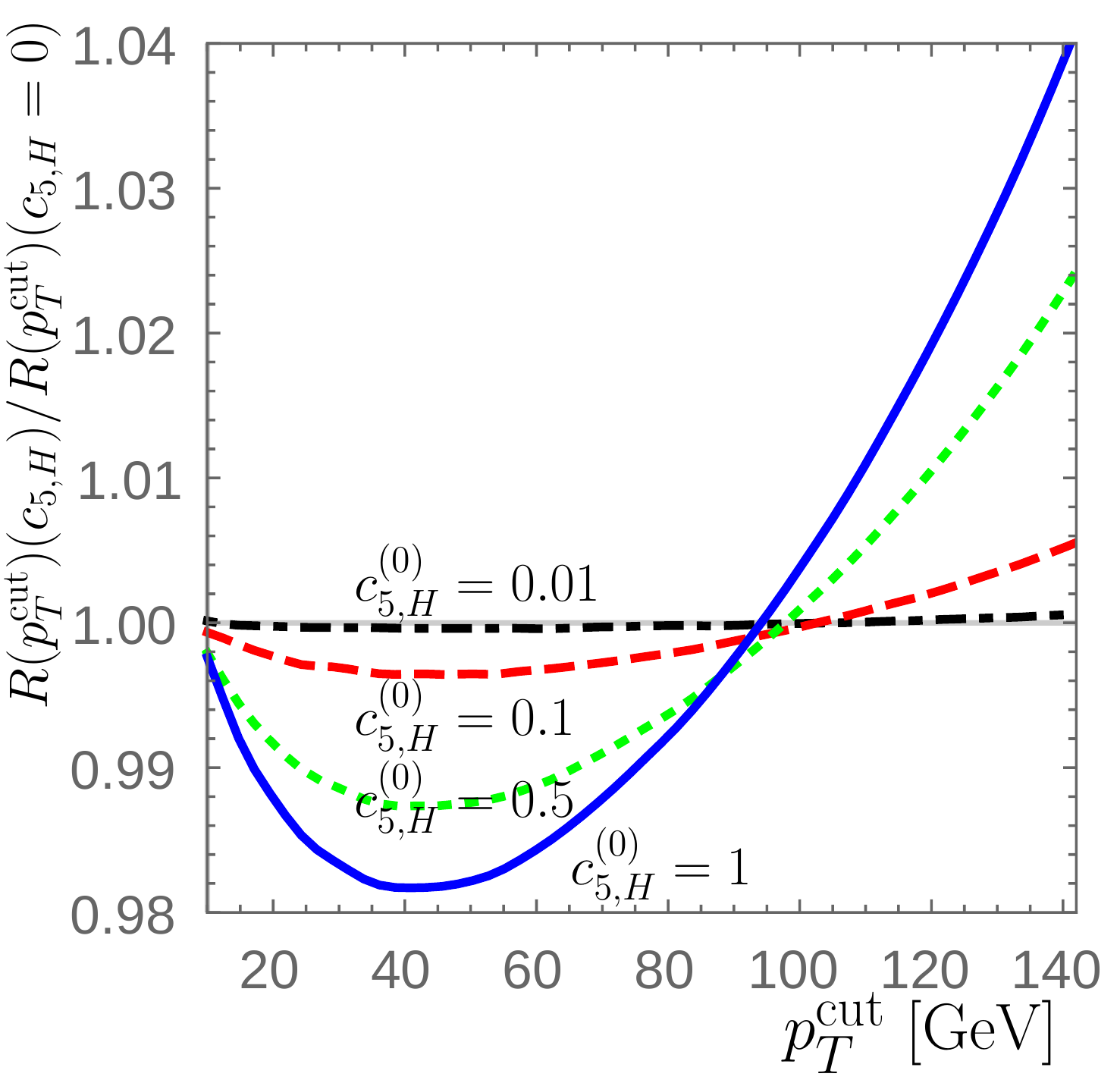} &
\includegraphics[width=0.47\textwidth]{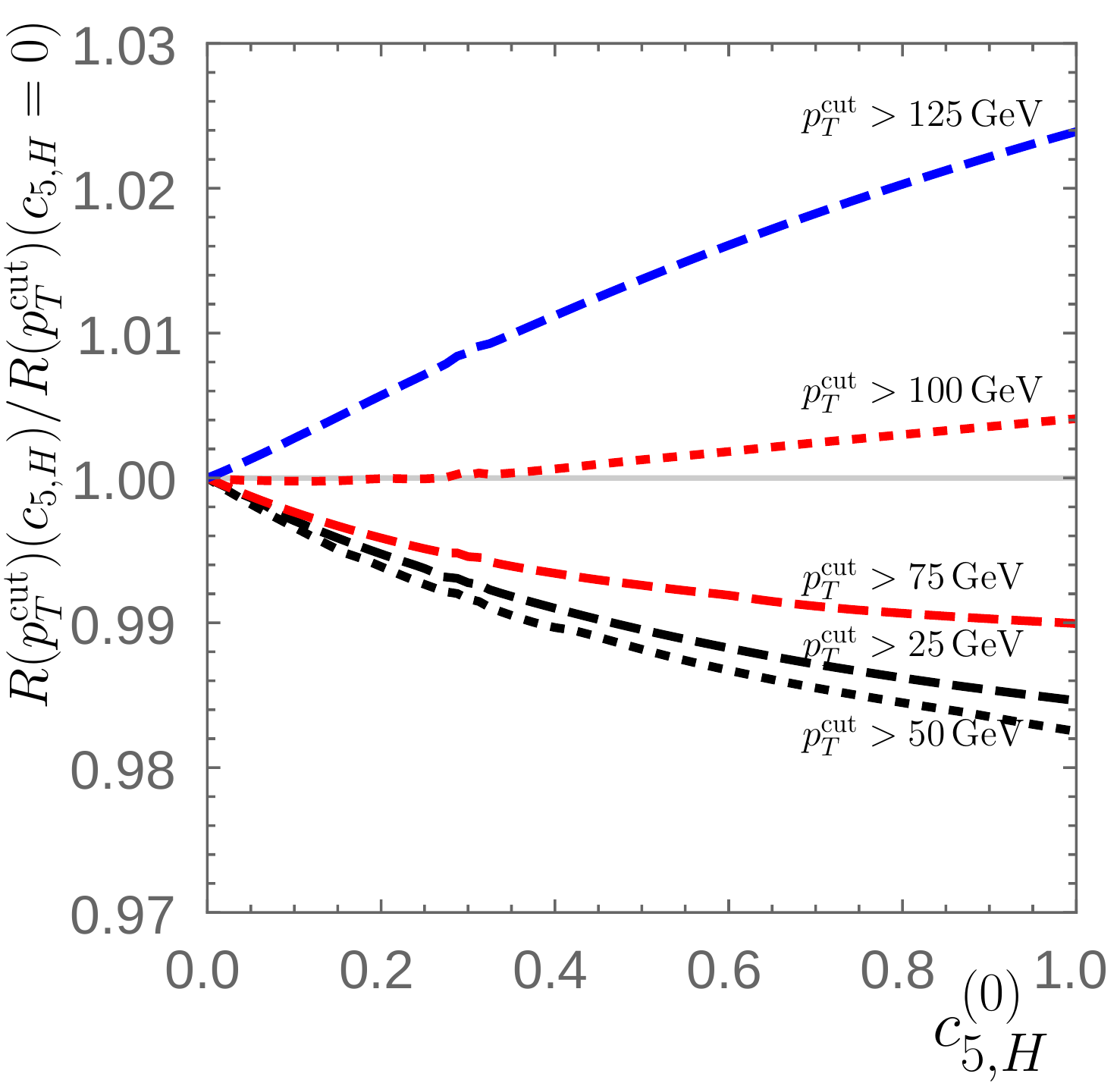}  \\[-0.4cm]
 (a) & (b)
\end{tabular}
\end{center}
\vspace{-0.7cm}
\caption{
(a) Ratio of $R(\pt{}^\text{cut})$ with different $c_{5,H}^{(0)}$ (see figure)
and $R(\pt{}^\text{cut})$ with $c_{5,H}=0$ as a function of $\pt{}^\text{cut}$ in GeV;
(b) Ratio of $R(\pt{}^\text{cut})$ and $R(\pt{}^\text{cut})$ with $c_{5,H}=0$
as a function of $c_{5,H}^{(0)}$ for different $\pt{}^\text{cut}$ (see figure).
In both figures we set $c_{5,A}^{(1)}=\tfrac{11}{4}c_{5,A}^{(0)}$.
Both figures are obtained for a \cp{}-even \sm{} Higgs
with $\mhiggs=125$\,GeV at the $\sqrt{s}=13$\,TeV \lhc{}.}
\label{fig:r-11}
\end{figure}

\begin{figure}
\begin{center}
\begin{tabular}{cc}
\includegraphics[width=0.47\textwidth]{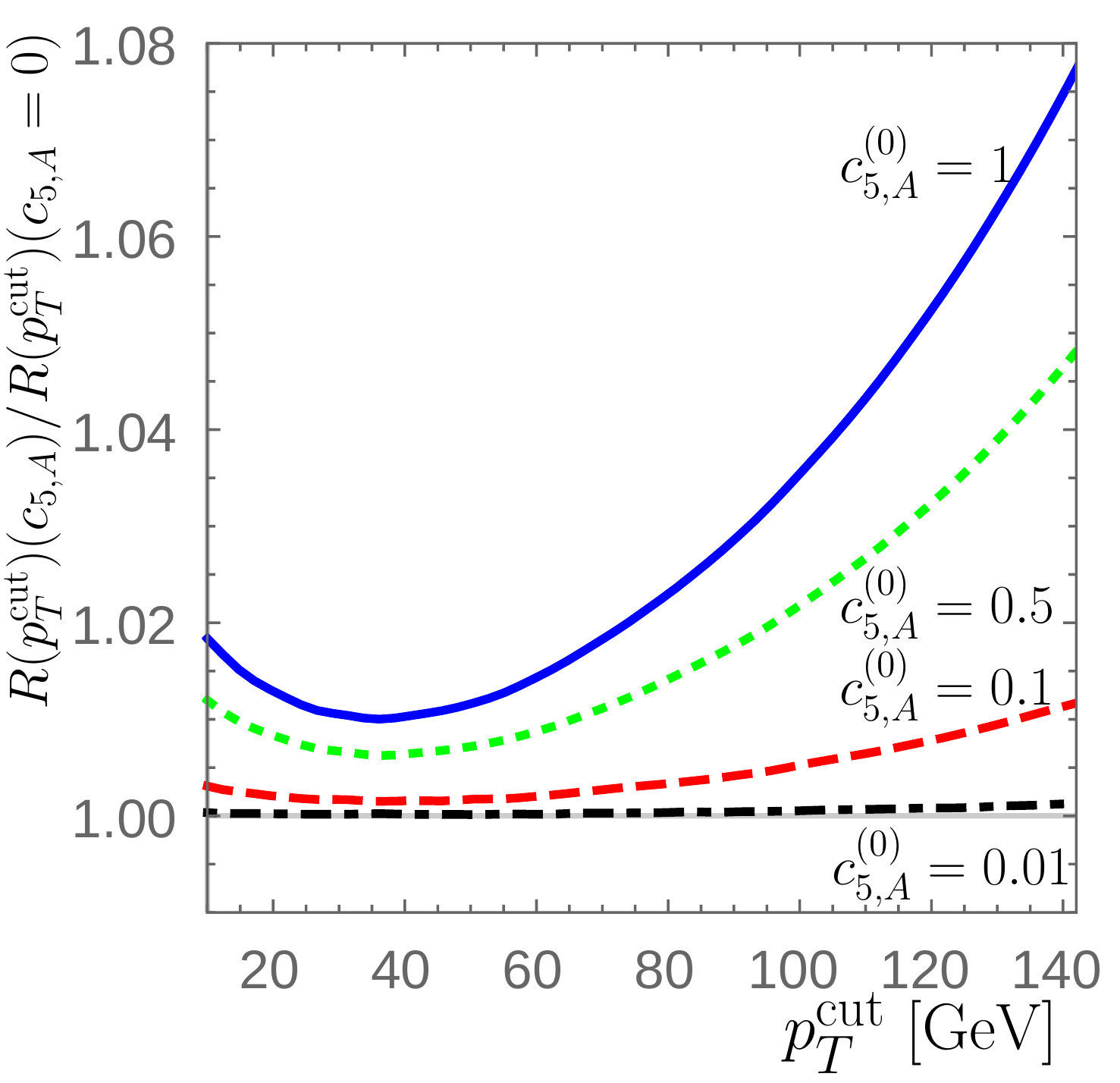} &
\includegraphics[width=0.47\textwidth]{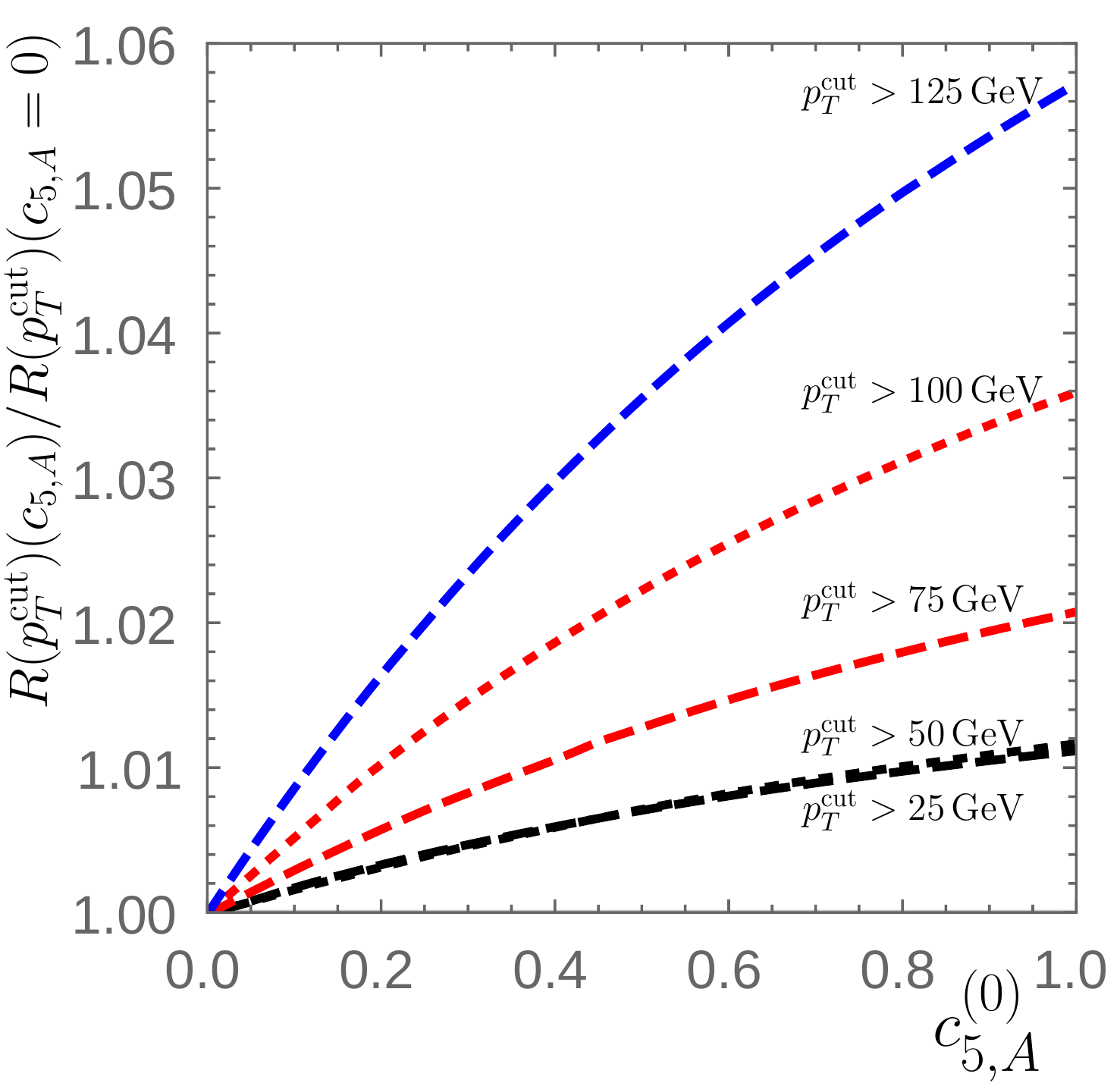}  \\[-0.4cm]
 (a) & (b)
\end{tabular}
\end{center}
\vspace{-0.7cm}
\caption{
(a) Ratio of $R(\pt{}^\text{cut})$ with different $c_{5,A}^{(0)}$ (see figure)
and $R(\pt{}^\text{cut})$ with $c_{5,A}=0$ as a function of $\pt{}^\text{cut}$ in GeV;
(b) Ratio of $R(\pt{}^\text{cut})$ and $R(\pt{}^\text{cut})$ with $c_{5,A}=0$
as a function of $c_{5,A}^{(0)}$ for different $\pt{}^\text{cut}$ (see figure).
In both figures we set $c_{5,A}^{(1)}=\tfrac{11}{4}c_{5,A}^{(0)}$.
Both figures are obtained for a \cp{}-odd Higgs
with $m_A=125$\,GeV at the $\sqrt{s}=13$\,TeV \lhc{}.
}
\label{fig:r-21}
\end{figure}

The ratio $R(\pt{}^\text{cut})$ of \eqn{eq:ptrat} is shown in
\fig{fig:r-11} for the \sm{} Higgs boson as a function of
(a)~$\pt{}^\text{cut}$ for various values of $c_{5,H}^{(0)}$, and (b)~$c_{5,H}^{(0)}$
for various values of $\pt{}^\text{cut}$.  Similarly, \fig{fig:r-21}
shows the ratio for a \cp{}-odd Higgs boson with mass $125$\,GeV. For
\fig{fig:r-11}~(a) and \fig{fig:r-21}~(a), $\sigma^\text{tot}$ is chosen
such that each $R(\pt{}^{\text{cut}})$ is normalized to its \nlo{}
inclusive cross section.  For \fig{fig:r-11}~(b) and \fig{fig:r-21}~(b),
$\sigma^{\text{tot}}=\sigma(\pt{}^{\text{cut}})$ for $c_5=0$ to ensure
that all curves start at one.  The minima, which are clearly visible
around $\pt{}^{\text{cut}}=50$\,GeV, are induced by the negative
interference with the bottom-quark induced contributions to gluon
fusion, which turns into a positive interference for higher values of
$\pt{}^{\text{cut}}$. Accordingly, these minima affect also the
dependence on $c_5$ in \fig{fig:r-11}~(b) and
\fig{fig:r-21}~(b), i.e.\ the lowest curve is obtained for a value of $\pt{}^{\text{cut}}$
around $40$\,GeV. Apart from the impact on the
inclusive cross section, the point-like
interaction encoded in the coefficient $c_5$ thus distorts the shape of
the $\pt{}$ distributions with respect to the loop-induced massive top-
and bottom-quark contributions, as expected.

Following the study performed in \citere{Grojean:2013nya}, we now work
out the dependence of the cross section with a minimal cut on $\pt{}$ on
the factors $\kappa_t$ and $c_{5,H}^{(0)}$ for the \sm{} Higgs
boson\footnote{Our $c_{5,H}^{(0)}$ corresponds to $\kappa_g$ in
\citere{Grojean:2013nya}.}.  In addition, we include the dependence on
the bottom-quark induced contribution through the factor $\kappa_b$,
since the latter is non-negligible for $\pt{}^\text{cut}<200$\,GeV.  For
this study we also choose $\pt{}$-dependent renormalization and
factorization scales $\muR=\muF=\sqrt{\mhiggs^2+\pt{}^2}/2$, which is
possible through the setting \blockentry{SCALES}{3}{=1}.  We define
$\tilde{\sigma}(\pt{}^\text{cut})$, which just includes the top-quark
induced contribution, i.e.\ we set $\kappa_t=1$ and $c_{5,H}=\kappa_b=0$,
and then perform a fit of
\begin{equation}
 \frac{\sigma(\pt{}^\text{cut})}{\tilde{\sigma}(\pt{}^\text{cut})}
 =(\kappa_t+c_{5,H}^{(0)})^2+\delta\kappa_tc_{5,H}^{(0)}+\epsilon (c_{5,H}^{(0)})^2
 +\delta_{bt}\kappa_b\kappa_t+\delta_{bg}\kappa_bc_{5,H}^{(0)}+\epsilon_b\kappa_b^2\,, 
\end{equation}
where we set $c_{5,H}^{(1)}=\tfrac{11}{4}c_{5,H}^{(0)}$ and
$\delta$ and $\epsilon$ are defined identically to
\citere{Grojean:2013nya}.
In addition, however, we include the
bottom-quark induced contribution, which is understood as pure
correction entering through $\delta_{bg}$, $\delta_{bt}$, and
$\epsilon_{b}$.  The values for $\delta$ and $\epsilon$ coincide at the
percent level with the values of Table~$1$ in \citere{Grojean:2013nya},
where for completeness we note that our calculation also includes the
$qq$ induced contribution to gluon fusion. For our numerical setup we
show the dependence of the five correction factors on the lower cut
$\pt{}^\text{cut}$ in \fig{fig:corrfactors}.

As can be seen in \fig{fig:corrfactors}~(a), the larger the lower cut
$\pt{}^\text{cut}$, the more the degeneracy between $\kappa_t$ and
$c_{5,H}$, which are indistinguishable in the inclusive cross section, is
broken.  On the other hand \fig{fig:corrfactors}~(b) points out that for
low $\pt{}^\text{cut}<200$\,GeV bottom-quark induced contributions
should also be taken into account.  The interferences of the latter with
the top-quark induced contributions on the one hand and with the
effective coupling $c_{5,H}$ on the other hand, encoded in $\delta_{bt}$
and $\delta_{bg}$, are identical only for low $\pt{}^\text{cut}$.  We
note that the cross section prediction for the \sm{} Higgs boson of
course should include the full correction by bottom quarks given by
$\delta_{bt}$ and $\epsilon_b$.  For completeness we partially also
reproduced Fig.~$2$ of \citere{Grojean:2013nya}, which illustrates the
disentanglement of the degeneracy between $\kappa_t$ and $c_{5,H}$.

As a last example we discuss the calculation of the gluon-fusion
cross section for an arbitrary scalar, which couples to gluons
through an effective operator $c_{5}^{(0)}=1$ only. Motivated
by the background deviation in the diphoton channel at $750$\,GeV
in both \lhc{} experiments~\cite{CMS:2016owr,ATLAS750}, we choose
the mass of the scalar to be $m_X=750$\,GeV. We pick an input
file for the \sm{}, set the \sm{} Higgs-boson mass to $\mhiggs{}=750$\,GeV,
include a \dimension{5} operator through \blockentry{DIM5}{11}{=1},
but set the \sm{} Higgs-boson couplings
to quarks and gauge bosons to zero in {\tt Block FACTORS} and 
through \blockentry{SUSHI}{7}{=0}. The results are shown
in \tab{tab:750}. We include the renormalization scale
uncertainty $\pm \Delta(\muR)$, which was obtained simultaneously. 
Again $\Delta(\muR)$ is the maximum
deviation of the cross section within the interval $\muR
\in[1/4,1]m_X$ and $\muR \in[1/2,2]m_X$ for
the central scale choices $\muR=\muF=m_X/2$ and
$\muR=\muF=m_X$, respectively. For this
purpose the Wilson coefficient is evolved perturbatively,
i.e. \blockentry{DIM5}{0}{=1}. At \nklo{3} the
soft expansion is performed up to $(1-x)^{16}$ with $a=0$.
The matching to the high-energy limit, $x\to 0$, is not applied.
Similar to the \sm{} Higgs boson we observe a good convergence of
the perturbative series with a renormalization scale
uncertainty of less than $\pm 1.3$ and $\pm 2.9$\% at \nklo{3} \qcd{}
for the central scale choices $\muR=\muF=m_X/2$ and $\muR=\muF=m_X$,
respectively.

\begin{table}[h]
\begin{center}
\begin{tabular}{|c|cc|}
\hline
$\sigma(gg\to X)$ [fb] & $\muR=\muF=m_X/2$ & $\muR=\muF=m_X$ \\
\hline
\lo{}      & $246.2  \pm 52.8$ & $185.8  \pm 36.0$\\
\nlo{}     & $368.7  \pm 43.1$ & $316.3  \pm 39.1$\\
\nnlo{}    & $410.0  \pm 19.1$ & $384.9  \pm 24.0$\\
\nklo{3}   & $414.6  \pm 5.4$  & $407.2  \pm 11.7$\\
\hline
\end{tabular}
\caption[]{\label{tab:750} Inclusive gluon-fusion cross section in fb
for a \cp{}-even scalar with mass $m_X=750$\,GeV, which couples to gluons through $c_{5}^{(0)}=1$ only.
The results are given at different orders \nklo{k}, $k=0,1,2,3$, in \qcd{}
for the $\sqrt{s}=13$\,TeV \lhc{} for two renormalization
and factorization scale choices.
The depicted uncertainty is the renormalization-scale uncertainty $\pm \Delta(\muR)$.}
\end{center}
\end{table}

\section{Conclusions}

We presented the new features implemented in version {\tt \sushiversion}
of the code \sushi{}.  Aside from the implementation of heavy-quark
annihilation, many new features aim at the improvement of the
gluon-fusion cross-section prediction and its associated uncertainty
estimate. In particular, \sushi{} now provides the soft expansion around
the threshold of Higgs production and the matching to the high-energy
limit for \cp{}-even Higgs bosons, at \nlo{}, \nnlo{} and
\nklo{3} \qcd{}. Top-quark mass effects beyond the usual infinite top-mass
limit can be taken into account at \nlo{} and \nnlo{}.  We investigated
the relevance of these effects for a \sm{}-like Higgs boson with a mass
of $125$\,GeV and provide a prediction of the corresponding gluon-fusion
cross section at the \lhc{} with a center-of-mass energy of $13$\,TeV.
Both for \cp{}-even and -odd Higgs bosons, \sushi{} now calculates the
renormalization-scale uncertainty simultaneously to the calculation of
the gluon-fusion cross section at the central scale.  Moreover, the
effects of \dimension{5} operators can be studied in any model currently
supported by \sushi{}.  We showed how the degeneracy between the
top-quark mass contribution and a point-like \dimension{5} operator
contribution can be broken at large values of the transverse momentum of
a Higgs boson with mass $125$\,GeV. The implementation of arbitrary
\dimension{5} operators is also particularly suited for the study of new
\cp{}-even and -odd scalars beyond the implemented models. We showed the
convergence of the perturbative series for the inclusive gluon-fusion
cross section of a scalar with mass $750$\,GeV at the $13$\,TeV \lhc{}.

Our description and the subsequent appendix include examples how the user can control the new features through
the setting of blocks in the input file of \sushi{}. Example input files are
contained in the {\tt example}-folder of the current \sushi{} release to be found
at \cite{sushiwebpage}.

\section*{Acknowledgments}

RVH would like to thank {\abbrev DFG} for financial support.  SL
acknowledges support by the {\abbrev SFB} 676 ``Particles, Strings and
the Early Universe''. Many of the calculations presented in this paper
have been performed on the {\abbrev FUGG} cluster at Bergische
Universit\"at Wuppertal.

\appendix
\section{Example input}
\label{app:inputfile}

In this appendix we present exemplary input blocks, which control
features that have been added to \sushi{} since its original release
described in \citere{Harlander:2012pb} (version {\tt 1.0.0}). This
includes the features described in the main text of this paper, but also
others like the introduction of the \nmssm{}\,\cite{Liebler:2015bka} and
the \thdm{}, or the calculation of the Higgs cross section in general
heavy-quark annihilation through \nnlo{}\,\cite{Harlander:2015xur}.  For
a complete up-to-date manual of \sushi{}, we refer the reader to
\citere{sushimanual}.

We begin with the main {\tt Block SUSHI} which may look as follows:
\begin{lstlisting}
Block SUSHI
    1  0  # Chosen model: 0=SM, 1=MSSM, 2=2HDM, 3=NMSSM
    2  11 # 11/12/13=scalar, 21/22=pseudo-scalar   
    3  0  # Particle collider: 0=pp, 1=ppbar
    4  1.3E+04   # center-of-mass energy in GeV
    5  3  # Order for ggh
    6  2  # Order for bbh
    7  2  # Electroweak contributions to ggh
   19  1  # 0 = silent mode of SusHi, 1 = normal output
   20  10 # ggh@nnlo subprocesses: 0=all, 10=ind. contributions
   21  0  # bbh@nnlo subprocesses: 0=all
\end{lstlisting}
\blockentry{SUSHI}{1}{} controls the physics model,
\blockentry{SUSHI}{2}{} the Higgs boson under consideration. In contrast
to the original release with the \thdm{}-adapted options
\blockentry{SUSHI}{2}{$\in$\{0,1,2\}} for the light, the pseudo-scalar
and the heavy Higgs boson, respectively, the Higgs bosons are now
defined in the more general \nmssm{}-framework, where
\blockentry{SUSHI}{2}{$\in\{$11,12,13$\}$} denote the \cp{}-even Higgs
bosons $\{H_1,H_2,H_3\}$, and \blockentry{SUSHI}{2}{$\in\{$21,22$\}$}
the \cp{}-odd Higgs bosons $\{A_1,A_2\}$. If \blockentry{SUSHI}{1}{=0},
$H_1$ assumes the role of the \sm{} Higgs boson, while $A_1$ is a
hypothetical pseudo-scalar with ``\sm-like'' couplings (see, e.g.,
\citere{Harlander:2002vv} for details). If
\blockentry{SUSHI}{1}{$\in\{$1,2$\}$}, then $H_1=h$ and $H_2=H$ are the
light and the heavy \cp-even Higgs boson, and $A_1=A$ is the \cp-odd
Higgs boson of the \mssm{} or the \thdm.

\blockentry{SUSHI}{5}{} now allows for the settings
\blockentry{SUSHI}{5}{=3}, activating the \nklo{3} \qcd{} top-quark
contributions as discussed in this paper, and \blockentry{SUSHI}{5}{=12}
for the inclusion of approximate \nnlo{} stop contributions as described
in \citere{Bagnaschi:2014zla} (see also \citere{Harlander:2003kf}).
Other new options are set in \blockentry{SUSHI}{$n$}{} with $n\geq 10$:
\blockentry{SUSHI}{19}{} controls the screen output verbosity of
\sushi{}, while \blockentry{SUSHI}{20}{} and \blockentry{SUSHI}{21}{}
allow to display the individual
subchannels in the calculations of gluon-fusion and bottom-quark
annihilation, respectively.

We continue with a description of the input blocks to control the
effects described in this paper. For this purpose we consider the input
file {\tt SM-N3LO\_best.in}, which contains new input entries in {\tt
  Block GGHMT}, {\tt Block GGHSOFT}, and {\tt Block SCALES}.  The {\tt
  Block GGHMT} controls top-quark mass effects in the calculation of the
gluon fusion cross section and has the following entries:
\begin{lstlisting}
Block GGHMT
   -1  3   # factor out exact LO result at LO(=0)/NLO(=1)/etc.
    0 -1   # expand through 1/mt^n at LO [-1=exact]
    1  4   # expand through 1/mt^n at NLO
   11  4   # expand gg through 1/mt^n at NLO
   12  4   # expand qg through 1/mt^n at NLO
   13  4   # expand qqbar through 1/mt^n at NLO
    2  4   # expand through 1/mt^n at NNLO
   21  4   # expand gg through 1/mt^n at NNLO
   22  4   # expand qg through 1/mt^n at NNLO
   23  4   # expand qqbar through 1/mt^n at NNLO
   24  4   # expand qq through 1/mt^n at NNLO
   25  4   # expand qqprime through 1/mt^n at NNLO
    3  0   # expand through 1/mt^n at N3LO (more not implemented)
   10  1   # [0/1]: do not/match to high energy limit at NLO
   20  1   # [0/1]: do not/match to high energy limit at NNLO
   30  1   # [0/1]: do not/match to high energy limit at N3LO
\end{lstlisting}
In this example, we factor out the the \lo{} top-quark mass dependence
through the first setting. The parameter \blockentry{GGHMT}{0}{} is
only of relevance for \blockentry{GGHMT}{-1}{=-1}
(meaning that the \lo{} top-quark mass dependence is not factored out)
and allows to control the top-quark mass terms in the \lo{} contribution in this case.
Beyond \lo{}, we take into account the
first four terms in the expansion in $1/\mtop$ in all channels.  Finally,
through \blockentry{GGHMT}{$n\cdot 10$}{=1} with $n\geq 1$, we match to
the high-energy limit at each order beyond \lo{}.

The block {\tt Block GGHSOFT} activates the soft expansion around the threshold
of Higgs production through \blockentry{GGHSOFT}{n,1}{=1} with $n\in\lbrace 1,2,3\rbrace$
and performs it up to $(1-x)^{16}$ with $a=0$, see \eqn{eq:softexp}:
\begin{lstlisting}
Block GGHSOFT # parameters for soft expansion
    1  1  16  0   # NLO  [0/1=n/y] [order] sig(x)/x^[n]
    2  1  16  0   # NNLO [0/1=n/y] [order] sig(x)/x^[n]
    3  1  16  0   # N3LO [0/1=n/y] [order] sig(x)/x^[n]
\end{lstlisting}

Let us next consider the new settings in {\tt Block SCALES}:
\begin{lstlisting}
Block SCALES
    1  5.00000000E-01   # Renormalization scale muR/mh for ggh
    2  5.00000000E-01   # Factorization scale muF/mh for ggh
  101  5.00E-01  2.00E+00   # min and max for muR uncertainty around SCALES(1)
  102  1.00E-01  1.00E+01   100   # min/max/steps for table of muR variation
   11  1.00000000E+00   # Renormalization scale muR/mh for bbh
   12  2.50000000E-01   # Factorization scale muF/mh for bbh  
\end{lstlisting}
Entry \blockentry{SCALES}{101}{} defines the interval of $\muR$ relative
to the central scale given in \blockentry{SCALES}{1}{}, which \sushi{}
uses to determine the renormalization scale uncertainty of the gluon
fusion cross section following \sct{sec:scaledep}.  Thus, in this
example, the $\muR$ uncertainty is obtained from a variation of $\muR$
within the interval $\muR\in[0.5,2]\cdot 0.5\cdot\mhiggs$.  The minimal
and maximal values of $\muR$ with respect to the central scale choice
relevant for the output file {\tt <outfile>\_murdep} are specified in
\blockentry{SCALES}{102}{}. In the example, the cross sections at 100
$\muR$ values within $[1/10,10]\cdot 0.5\cdot\mhiggs$ will be printed in
the additional output file. Finally, note that the renormalization and
factorization scales for bottom-quark annihilation can be set
independently in \blockentry{SCALES}{11}{} and (12), which is possible
since \sushi{} release {\tt 1.2.0}.

For the discussion of \dimension{5} operators we add
an example of an arbitrary scalar~$\phi$ coupling to gluons,
similar to the one in the last paragraph of \sct{sec:numdim5}.
\begin{lstlisting}
Block DIM5
    0  1   # Running of DIM5 operators 0=off/1=pert./2=res.
   11  1.00000000E+00   # LO coeff of dim-5 operator
  111  0.00000000E+00   # NLO coeff of dim-5 operator
  211  0.00000000E+00   # NNLO coeff of dim-5 operator
  311  0.00000000E+00   # N3LO coeff of dim-5 operator
\end{lstlisting}
We thus consider a \cp{}-even Higgs boson through
\blockentry{SUSHI}{2}{=11} and control its coupling to gluons through
\blockentry{DIM5}{$n\cdot$100$+$11}{} with $n\in\{0,1,2,3\}$.  The
setting \blockentry{DIM5}{0}{=1} evolves the \dimension{5} operator
perturbatively, and through \blockentry{DIM5}{11}{=1} we only fix the
lowest order contribution of the \dimension{5} operator at the
scale~$\mphi$ to a non-vanishing value.  Higher order contributions will
be generated through the perturbative running and can be taken from the
output {\tt Block DIM5OUT} in the output file.

Next we display an example of the calculation of the cross section
$c\bar c\to H$ in the \sm{} through the settings in {\tt Block QQH},
where apart from the incoming parton types also the Yukawa coupling in
GeV and the input scale of the Yukawa coupling in GeV have to be
specified:
\begin{lstlisting}
Block QQH
    1  4   # parton 1 = c   
    2 -4   # parton 2 = cbar
   11  1.27500000E+00   # Yukawa coupling
   12  1.27500000E+00   # renorm.-scale for input Yuk.-coupl.
\end{lstlisting}

Let us next discuss the access to new physics models within \sushi{}.
Since {\tt SusHi\_1.0.2}, calculations in the \cp{}-conserving \thdm{}
can be performed. Release {\tt 1.1.1} introduced a link to the external
code {\tt 2HDMC}~\cite{Eriksson:2009ws}.  The input of the \thdm{}
closely resembles the \sm{} input. The user has to specify the Higgs
under consideration using \blockentry{SUSHI}{2}{$\in\{$11,12,21$\}$} for the
light, the heavy, and the pseudo-scalar Higgs, respectively.
Moreover, the corresponding Higgs mass has to be given in {\tt Block
  MASS} together with the Higgs mixing angle~$\alpha$ in {\tt Block
  ALPHA} and the ratio of the vacuum expectation values $\tan\beta$ in
{\tt Block MINPAR}. The different types of the \thdm{} as discussed in
\citere{Branco:2011iw} can be distinguished in {\tt Block 2HDM}:
\begin{lstlisting}
Block 2HDM # 2HDM version according to Ref. [114]
    2  # (1=Type I,2=Type II,3=Flipped,4=Lepton Specific)
Block MINPAR
    3 5.00000000E+00  # tanb
Block ALPHA
   -5.00000000E-01    # mixing in CP-conserving Higgs sector
Block MASS
   25  1.25000000E+02 # Higgs mass h
   35  1.50000000E+02 # Higgs mass H
   36  3.00000000E+02 # Pseudoscalar Higgs mass A
\end{lstlisting}

The link to {\tt 2HDMC} allows the user to provide the \thdm{} input
parameters in {\tt Block 2HDMC} either in the $\lambda$ basis, the physical basis, or the H2
basis. For details we refer to the {\tt 2HDMC} manual.
{\tt Block 2HDMC} makes the blocks {\tt 2HDM, MINPAR, ALPHA } and {\tt MASS }
obsolete. Example input
files are provided in the {\tt example}-folder of the current \sushi{}
distribution.  For example, in case of the $\lambda$ basis, the relevant
block takes the form:
\begin{lstlisting}
Block 2HDMC	# 2HDMC (Ref. [88])
    1  1            # 2HDMC key, 1=lambda basis, 2=physical basis, 3=H2 basis
    2  2               # 2HDM version type
    3  1.00000000E+01  # tan(beta)
    4  1.00000000E+02  # m12
   11  1.00000000E-01  # lambda1
   12  2.00000000E-01  # lambda2
   13  3.00000000E-01  # lambda3
   14  4.00000000E-01  # lambda4
   15  5.00000000E-01  # lambda5
   16  0.00000000E-01  # lambda6
   17  0.00000000E-01  # lambda7
\end{lstlisting}

Since version {\tt 1.5.0}, \sushi{} includes also the
\nmssm{}\,\cite{Liebler:2016dpn}.  For the corresponding input blocks we
refer to \citere{Liebler:2015bka} (note, however, a change in the
convention for {\tt Block NMAMIX}~\cite{Liebler:2016dpn}).  Contrary to
the \mssm{} and the \thdm{}, no link to an external code is provided,
such that the \nmssm{} Higgs sector has to be specified completely in
the input file. This involves the Higgs masses in {\tt Block MASS} as
well as the Higgs mixing in the \cp{}-even sector in {\tt Block NMHMIX}
and the \cp{}-odd sector in {\tt Block NMAMIX}.

{\tt SusHi\_1.5.0} also introduced the {\tt Block FEYNHIGGSFLAGS}, which
allows to control the various options of a {\tt
  FeynHiggs}~\cite{Heinemeyer:1998yj,Heinemeyer:1998np,Degrassi:2002fi,Frank:2006yh}
run in the \mssm{}.  The number of arguments depends on the {\tt
  FeynHiggs} version, please consider the \mssm{} input and output files
in the {\tt example}-folder of each \sushi{} release.  We generally
provide input files for all models and links to external codes in the
folder {\tt example}, which is part of each \sushi{} release. We thus
encourage the user to start his/her considerations with an {\tt
  example} input file.

\end{document}